\documentclass[aps, prd, 10pt, twocolumn, superscriptaddress, nofootinbib, floatfix]{revtex4-2}

\usepackage{graphicx}
\usepackage{amsmath}
\usepackage{amssymb}
\usepackage{lmodern}
\usepackage{footnote}
\usepackage{booktabs}
\usepackage[colorlinks=false, pdfborder={0 0 0}]{hyperref}

\renewcommand{\vec}[1]{{\bf #1}}
\newcommand{\grad}{\vec{\nabla}}
\newcommand{\hub}{\mathcal{H}}
\newcommand{\mpl}{M_{\rm Pl}}
\newcommand{\mphi}{M_{\phi}}
\newcommand{\gcs}{g_{\rm cs}}
\newcommand{\mpc}{{\rm Mpc}}
\newcommand{\kp}{k_{\rm p}}
\newcommand{\As}{A_{\rm s}}
\newcommand{\ns}{n_{\rm s}}
\newcommand{\Lbox}{L_{\rm box}}
\newcommand{\Vbox}{V_{\rm box}}
\newcommand{\Ngrid}{N_{\rm grid}}

\newcommand{\dee}{{\rm d}}

\newcommand{\dd}{\delta_{\rm D}^{(3)}}

\usepackage{xcolor}

\begin{document}
	
\title{Primordial Power Spectrum and Bispectrum from Lattice Simulations of Axion-U(1) Inflation}

\author{Drew Jamieson}
\email{jamieson@mpa-garching.mpg.de}
\affiliation{Max-Planck-Institut f\"ur Astrophysik, Karl-Schwarzschild-Straße 1, 85748 Garching, Germany}

\author{Angelo Caravano}
\email{caravano@iap.fr}
\affiliation{Institut d'Astrophysique de Paris, UMR 7095 du CNRS et de Sorbonne Universit\'e, 98 bis Bd Arago, 75014 Paris, France}

\author{Eiichiro Komatsu}
\affiliation{Max-Planck-Institut f\"ur Astrophysik, Karl-Schwarzschild-Straße 1, 85748 Garching, Germany}
\affiliation{
Ludwig-Maximilians-Universit\"at M\"unchen, Schellingstr. 4, 80799 M\"unchen, Germany
}
\affiliation{Kavli Institute for the Physics and Mathematics of the Universe (Kavli IPMU, WPI), Todai Institutes for Advanced Study, The University of Tokyo, Kashiwa 277-8583, Japan}

\date{\today}
	
\begin{abstract}
    We present primordial non-Gaussianity predictions from a new high-precision code for simulating axion-U(1) inflation on a discrete lattice. We measure the primordial scalar curvature power spectrum and bispectrum from our simulations, determining their dependence on both scale and axion-gauge coupling strength. Both the gauge-sourced power spectrum and the bispectrum exhibit a strong blue tilt due to our choice of an $\alpha$-attractor inflaton potential. We provide fitting functions for the power spectrum and bispectrum that accurately reproduce these statistics across a wide range of scales and coupling strengths. While our fitting function for the bispectrum has a separable form, results from high-resolution simulations demonstrate that the full shape is not separable. Thus, our simulations generate realizations of primordial curvature perturbations with nontrivial correlators that cannot be generated using standard techniques for primordial non-Gaussianity. We derive bounds on the axion-gauge coupling strength based on the bispectrum constraints from the cosmic microwave background, demonstrating a new method for constraining inflationary primordial non-Gaussianity by simulating the nonlinear dynamics.
\end{abstract}
	 
\maketitle

\section{Introduction}
\label{sec:intro}

Prospects for discovering new physics from the early universe face two encouraging opportunities. The first is an increase in the abundance and quality of observational data \cite{CMB-S4:2022ght,LiteBIRD:2022cnt,SimonsObservatory:2025wwn,LSST:2008ijt,EUCLID:2011zbd,PFSTeam:2012fqu, SPHEREx:2014bgr,Spergel:2015sza,DESI:2016fyo}, yielding high levels of statistical precision. The second is an ever-broadening range of theoretical models to guide our searches. Observational precision demands stringent systematic precision in our theoretical predictions. Meanwhile, theorists have expanded the phenomenology of early-universe cosmic inflation \cite{Starobinsky:1980te,Guth:1980zm,Sato:1981qmu,Albrecht:1982wi,Linde:1981mu}---the leading hypothesis explaining the origin of the universe---developing new classes of models with significant nonlinear interactions. Obtaining robust observational predictions for such models requires simulating their nonlinear dynamics.

Inflationary models featuring axion-like fields coupled to gauge fields exemplify these challenges. In axion-U(1) inflation, the inflaton is an axion-like field with a Chern-Simons coupling to a U(1) gauge field \cite{Anber:2009ua,Pajer:2013fsa}. Nonlinear interactions in this model generate an observable hierarchy of primordial non-Gaussian correlators \cite{Barnaby:2010vf,Barnaby:2011vw,Anber:2012du}, including a parity-violating scalar four-point function \cite{Niu:2022fki,Fujita:2023inz}, along with chiral gravitational waves \cite{Sorbo:2011rz,Anber:2012du}. 

The axion-U(1) model also exhibits strong backreaction and nonperturbative regimes \cite{Cheng:2015oqa,Ferreira:2015omg,Gorbar:2021rlt,Peloso:2022ovc,vonEckardstein:2023gwk,Galanti:2024jhw}, where the
simplified assumptions of semi-analytic methods break down. Computing even the duration of inflation in such models requires treating the full nonlinearity of their dynamics. In such regimes, direct numerical simulation becomes indispensable \cite{Cuissa:2018oiw,Caravano:2021bfn,Caravano:2022epk,Caravano:2022yyv,Figueroa:2023oxc,Figueroa:2024rkr,Caravano:2024xsb,Sharma:2024nfu,Iarygina:2025ncl}.

Lattice simulations have emerged as a powerful computational tool for studying the physics of inflation \cite{Prokopec:1996rr,Felder:2000hq,Bond:2009xx,Cuissa:2018oiw,Figueroa:2021yhd,Caravano:2021pgc,Caravano:2024tlp,Caravano:2024moy,Caravano:2025diq,Caravano:2025klk}. Such simulations are essential for accurately capturing a wider range of nonlinear dynamics that lie beyond the reach of perturbative, semi-analytic methods. In this work, we present results from a new code specifically designed to simulate axion-gauge inflation. We focus on the weak backreaction regime of this model, where the gauge field does not affect the background evolution. We solve the classical equations of motion on a discrete lattice in an expanding universe using pseudospectral methods combined with temporal grid refinement, high-order Runge-Kutta integration, and dynamical time-stepping. Our new simulation techniques achieve unprecedented precision in simulations of the early universe. This precision enables us to characterize the primordial signal in detail, including the full shape of the bispectrum. This level of detail and accuracy will be vital for obtaining robust constraints from upcoming survey data.

As we will demonstrate, simulating the nonlinear inflationary dynamics provides a method for generating realizations of the primordial density field with nontrivial, nonseparable forms of primordial non-Gaussianity. This allows us to overcome the inherent limitations of perturbative analyses and template-based searches, widening the class of testable inflationary models.

The paper is structured as follows. In Section \ref{sec:model}, we review the physics of the axion-U(1) model. Section \ref{sec:sim} details the simulation algorithm. In Section \ref{sec:res}, we present the power spectra and bispectra measured from our simulations, along with fitting functions motivated by perturbation theory that accurately capture the scale dependence and coupling strength dependence of these observables. We also provide an estimate of current upper bounds on the axion-gauge coupling strength based on existing constraints on the bispectrum of the cosmic microwave background (CMB). We conclude in Section \ref{sec:con} with an outlook on future applications of the simulation techniques developed and demonstrated in this work.

\section{Axion-U(1) Inflation}
\label{sec:model}

    We consider a model of inflation in which the inflaton $\phi$ is an axion-like field coupled to a U(1) gauge field $A_\mu$ via a Chern-Simons term. The action describing this system is \cite{Anber:2009ua}
    \begin{align}
        \label{eq:action}
        \begin{split}
            S = \int \mathrm{d}^4 x \sqrt{-g} \Biggl[ &\frac{\mpl^2}{2}R - \frac{1}{2}\partial_{\mu} \phi \partial^{\mu} \phi - V(\phi) \\
            &\quad - \frac{1}{4} F_{\mu\nu} F^{\mu\nu} - \frac{\gcs}{4} \phi F_{\mu\nu} \tilde{F}^{\mu\nu} \Biggr] \,,
        \end{split}
    \end{align}
    where $\mpl^2=(8\pi G)^{-1}$, and we set $c\equiv 1$ and $\hbar\equiv 1$. The parameter $\gcs$ represents the dimensionful axion-gauge coupling strength, with dimensions of inverse mass. Throughout this work, we focus on the dynamics of the scalar and gauge fields, neglecting the direct influence of metric perturbations except for their effects on the linearized inflaton equation of motion. We adopt a flat, inflationary FLRW cosmology with the metric $g_{\mu\nu}(\tau) = a(\tau)^2 \eta_{\mu\nu}$, where $\tau$ is the conformal time coordinate. We denote conformal time derivatives with primes.
	
    To analyze the gauge field dynamics, we define the potential $A_\mu = (-\varphi, \vec{A})$, where bold symbols denote comoving 3-vectors. The gauge field strength tensor is given by $F_{\mu\nu} = \partial_{\mu}A_{\nu} - \partial_{\nu}A_{\mu}$. We introduce the comoving electric and magnetic fields,
	\begin{align}
            \label{eq:Edef}
		\vec{E} &= -\vec{\grad} \varphi - \vec{A}' \,, \\
            \label{eq:Bdef}
		\vec{B} &= \grad\times\vec{A} \,,
	\end{align}
	which correspond to the components of the field strength tensor as $F_{0i} = -E_i$ and $F_{ij} = \epsilon_{ijk} B^k$. We treat $\vec{E}$ and $\vec{B}$ as comoving 3-vectors under the 3+1 metric decomposition, meaning spatial indices are raised and lowered by the Euclidean metric $\delta_{ij}$, and $\epsilon_{ijk}$ is the Levi-Civita symbol for the standard Euclidean three-dimensional cross product.
	
    The dual field strength tensor is defined as
    \begin{align}
        \tilde{F}^{\mu\nu} = \frac{1}{2} \epsilon^{\mu\nu\rho\lambda}F_{\rho\lambda} \,,
    \end{align}
    where $\epsilon^{\mu\nu\rho\lambda}$ is the Levi-Civita tensor with $\epsilon_{0123}=\sqrt{-g}$, implying $\epsilon^{0123}=-(\sqrt{-g})^{-1}$. The components of the dual field strength tensor can then be expressed as
    \begin{align}
            \label{eq:Bdual}
        \sqrt{-g} \tilde{F}^{0i} &= -B^i \,, \\
            \label{eq:Edual}
        \sqrt{-g} \tilde{F}^{ij} &= \epsilon^{ijk} E_k \,.
    \end{align}
    Therefore, $F_{\mu\nu}\tilde{F}^{\mu\nu}=+4\vec{E}\cdot\vec{B}$.
	
    The gauge field equations of motion, in their Maxwell equations form, are
    \begin{align}
        \label{eq:coulomb1}
        \grad \cdot \vec{E} &= \gcs \grad\phi \cdot \vec{B} \,, \\
        \label{eq:maxamp1}
        \vec{E}' - \grad \times \vec{B} &= -\gcs\Bigl(\phi'\vec{B} + \grad\phi\times\vec{E}\Bigr) \,.
    \end{align}
    By working in the comoving Lorenz gauge, $\partial_\mu A^{\mu} = 0$, we can solve the Coulomb constraint, eliminating the scalar gauge potential $\varphi$ and the longitudinal part of the vector potential. This leaves only the two transversal vector modes, which can be expressed as right- and left-helicity polarizations (see Appendix \ref{app:geom} for further details). The field equations for the Fourier modes of these polarizations satisfy
    \begin{align}
        \label{eq:ARL}
        A''_{\rm R/L}(\tau, \vec{k}) = -k^2\left(1 \mp \frac{2\hub\xi}{k}\right) A_{\rm R/L}(\tau, \vec{k}) + S^{\vec{A}}_{\rm R/L}(\tau, \vec{k}) \,.
    \end{align}
    Here, Fourier modes are distinguished from coordinate space fields by their argument unless stated otherwise. We have introduced the conformal Hubble rate, $\hub(\tau)=a'(\tau)/a(\tau)$, and the time-dependent parameter $\xi(\tau)$ controlling the coupling between the gauge modes and the inflationary background,
    \begin{align}
        \xi(\tau) = \frac{\gcs \bar{\phi}'(\tau)}{2\hub(\tau)} \,.
    \end{align}
    In coordinate space, the nonlinear source term is given by
    \begin{align}
        \label{eq:sARL}
        \vec{S}^{\vec{A}}(\tau, \vec{x}) = -\gcs \Bigl( \delta\phi'(\tau, \vec{x}) \vec{B}(\tau, \vec{x}) + \grad \phi(\tau, \vec{x})\! \times\! \vec{E} (\tau, \vec{x}) \Bigr) \,. 
    \end{align}
    We have separated the inflaton field into its background and fluctuating components, $\phi(\tau, \vec{x}) = \bar{\phi}(\tau) + \delta\phi(\tau, \vec{x})$.
	
    The inflaton field equation is
	\begin{align}
		\label{eq:inf1}
		-\square \phi = -2\hub\phi' -a^2\Delta m_{\rm eff}^2 \delta\phi - a^2 V_{,\phi} - \gcs a^{-2}\,\vec{E}\cdot\vec{B} \,,
	\end{align}
    where $-\square f = f'' - \grad^2 f$ is the comoving wave operator. The second term on the right-hand side represents a time-dependent effective mass shift resulting from integrating out the leading scalar metric perturbations (see Appendix \ref{app:seom}),
    \begin{align}
        \label{eq:mass_shift}
        a^2\Delta m_{\rm eff}^2 = 2\left(\frac{a''}{a} - \hub^2\right)\left(3 + 2\frac{\bar{\phi}''}{\hub\bar{\phi}'} - \frac{a''}{a}\right) \,.
    \end{align}
    The Fourier modes of the inflaton field satisfy
    \begin{align}
        \label{eq:phi2}
        \begin{split}
        \delta\phi''(\tau, \vec{k}) &= -\bigl(k^2 + a^2m_{\rm eff}^2\bigr)\delta\phi(\tau, \vec{k}) - 2\hub\delta\phi'(\tau, \vec{k}) \\
        &\quad + S^{\phi}(\tau, \vec{k}) \,,
        \end{split}
    \end{align}
    where the total time-dependent effective mass in this equation is
    \begin{align}
        m_{\rm eff}^2(\tau) = V_{,\phi\phi}\bigl(\bar{\phi}(\tau)\bigr) + \Delta m_{\rm eff}^2(\tau) \,,
    \end{align}
    and the nonlinear source term is computed locally in coordinate space,
    \begin{align}
        \label{eq:sphi}
        \begin{split}
        S^{\phi}(\tau, \vec{x}) &=
        - a^2 \Bigl(V_{,\phi}(\phi) - \langle V_{,\phi}\rangle - V_{,\phi\phi}(\bar{\phi})\delta\phi\Bigr) \\
        & \quad - \gcs a^{-2} \Bigl(\vec{E}\cdot\vec{B} - \langle \vec{E}\cdot\vec{B} \rangle \Bigr) \,.
        \end{split}
    \end{align}
    Angled brackets denote the average over all space. We have subtracted the background quantities and the linearized term from the inflaton potential, retaining only the nonlinear interactions.
	
    The approach delineated in this section, where we have integrated out linear metric perturbations while retaining all nonlinearities in the field dynamics, allows us to solve for the fully nonlinear evolution of the axion-gauge system while linearizing gravitational interactions.\footnote{This approach is analogous to the one followed in \cite{Caravano:2024xsb}.} This is valid as long as we remain in slow-roll inflation, where gravitational interactions are suppressed. The primary goal of this work is to simulate the physics of this system by solving Eqs.~\eqref{eq:ARL}, \eqref{eq:sARL}, \eqref{eq:phi2}, and \eqref{eq:sphi} on a discrete lattice with Gaussian random initial conditions, which represent the Bunch-Davies vacuum \cite{Bunch:1978yq} in the distant past. Before delving into the details of these simulations, we first review the phenomenology of this model.

\subsection{Gauge Field Production and Non-Gaussianity}
	
    The coupling between the inflaton background and the gauge field directly affects the linear gauge field mode functions. We can see this by setting the nonlinear source term in Eq.~\eqref{eq:ARL} to zero. The different signs for the right and left polarizations in the second term in the parentheses of Eq.~\eqref{eq:ARL} indicate that parity is dynamically violated \cite{Anber:2009ua}. Depending on the signs of the coupling strength and the background inflaton time derivative, one of the two polarizations is exponentially enhanced when \cite{Barnaby:2011vw}
    \begin{align}
        \label{eq:xi_bounds}
        \frac{1}{8|\xi(\tau)|} < \frac{k}{\hub(\tau)} < 2 |\xi(\tau)| \,,
    \end{align}
    whereas the other polarization is mildly suppressed. The exponential enhancement of one helicity state over the other signifies near-maximal parity violation in the gauge field. We will assume $\gcs>0$, which means the left-handed modes are enhanced when $\bar{\phi}'<0$.
	
    The linear-order gauge field enhancement imprints on the inflaton fluctuations at quadratic order through the $\vec{E}\cdot\vec{B}$ contribution to the nonlinear source term in Eq.~\eqref{eq:sphi}. This process is an inverse decay mechanism, where two inflaton-background-enhanced photons annihilate each other, producing an inflaton via the Chern-Simons interaction \cite{Barnaby:2010vf}. 
    Separating the inflaton fluctuations into vacuum and sourced perturbations:
    \begin{align}
        \label{eq:vac_src}
        \delta\phi(\tau, \vec{k}) = \delta\phi_{\rm vac}(\tau, \vec{k}) + \delta\phi_{\rm src}(\tau, \vec{k}) \,,
    \end{align}
    where the vacuum fluctuations are the linear, homogeneous solution to Eq.~\eqref{eq:phi2}, setting the nonlinear sources to zero. The sourced fluctuations are the particular solution to Eq.~\eqref{eq:phi2}, including the nonlinear sources. Both the inflaton potential and gauge field coupling contribute sourced fluctuations. However, in the model considered here, the gauge field fluctuations dominate due to their exponential enhancement.
	
    Outside the horizon, the helical gauge modes are no longer supported and decay away through redshift. Their influence is retained only through the sourced inflaton perturbations, which imprint on the primordial curvature perturbations that are conserved outside the horizon \cite{Weinberg:2003sw}. In the linear approximation, the curvature perturbation is given by \cite{Starobinsky:1982ee,Hawking:1982cz,Guth:1982ec,Bardeen:1983qw}
    \begin{align}
        \label{eq:zeta}
        \zeta(\vec{k}) = -\frac{\hub(\tau)}{\bar{\phi}'(\tau)} \delta\phi(\tau, \vec{k}) \bigg|_{-k\tau \ll 1} \,.
    \end{align}
    The vacuum and sourced parts of the inflaton field produce vacuum ($\zeta_{\rm vac}$) and sourced ($\zeta_{\rm src}$) primordial curvature perturbations.

    The gauge field and inflaton have independent, uncorrelated linear vacuum fluctuations. The leading order sourced inflaton fluctuations are quadratic in the gauge field vacuum fluctuations. Thus, the vacuum inflaton perturbations have negligible correlation with the sourced inflaton perturbations, and the primordial power spectrum has two contributions:
    \begin{align}
        \label{eq:pspec_def}
        \bigl\langle \zeta(\vec{k}) \zeta(\vec{k}') \bigr\rangle &\simeq 
        \bigl\langle \zeta_{\rm vac}(\vec{k}) \zeta_{\rm vac}(\vec{k}') \bigr\rangle + \bigl\langle \zeta_{\rm src}(\vec{k}) \zeta_{\rm src}(\vec{k}') \bigr\rangle \nonumber \\
        &= (2\pi)^3\delta(\vec{k}+\vec{k}')\bigl(P_{\rm vac}(k) + P_{\rm src}(k) \bigr) \,.
    \end{align}
    Introducing the dimensionless power spectrum,
    \begin{align}
        \mathcal{P}(k) \equiv \frac{k^3}{2\pi^2} P(k) \,,
    \end{align}
    the vacuum power spectrum is parameterized as
    \begin{align}
        \label{eq:pvac}
        \mathcal{P}_{\rm vac}(k) = \As \left(\frac{k}{\kp}\right)^{\ns-1} \,,
    \end{align}
	where $\kp = 0.05~{\rm Mpc}^{-1}$ is the pivot wave number.
    
    Assuming $\xi(\tau)$ is constant, one can analytically estimate the sourced part of the power spectrum arising from the nonlinear axion-gauge coupling. While we do not reproduce the full calculation here, the resulting expression for superhorizon scales is given by \cite{Anber:2009ua,Barnaby:2010vf,Barnaby:2011vw,Anber:2012du},
    \begin{equation}
        \label{eq:ps_th}
        \mathcal{P}_{\rm src}(k) = \As \mathcal{P}_{\rm vac}(k) e^{4\pi|\xi|} f_2(\xi)\,,
    \end{equation}
    where $f_2(\xi)$ is a function computed in Ref.~\cite{Barnaby:2011vw}. In the range $2 < |\xi| < 3$, corresponding to weak but observationally interesting production of primordial non-Gaussianity, $f_2(\xi)$ is approximated by
    \begin{equation}
        f_2(\xi) \simeq \frac{3 \times 10^{-5}}{|\xi|^{5.4}}.
    \end{equation}
    
    These results suggest that the only effect of the axion-gauge coupling on the power spectrum is an increase in the overall scalar amplitude. However, if $\xi(\tau)$ is not constant, its time variation translates into a scale dependence of the sourced power spectrum, thereby altering the shape of the total primordial power spectrum \cite{Anber:2009ua}. With increasing $|\xi(\tau)|$, the sourced power spectrum becomes blue-tilted and can dominate on small scales \cite{McDonough:2016xvu}. Since this effect leaves the large-scale power unaffected, it is distinct from the typical running of the spectral tilt. It would manifest as an elevation of small-scale primordial power, with the standard power spectrum given by Eq.~\eqref{eq:pvac} appearing on larger scales.
	
    The sourced curvature perturbations also carry primordial non-Gaussianity. The vacuum and sourced fluctuations are nearly uncorrelated, and the sourced field dominates the bispectrum:
    \begin{align}
        \label{eq:bspec_def}
        \begin{split}
        \bigl\langle \zeta(\vec{k}_1) \zeta(\vec{k}_2) \zeta(\vec{k}_3) \bigr\rangle &\simeq
        \bigl\langle \zeta_{\rm src}(\vec{k}_1) \zeta_{\rm src}(\vec{k}_2) \zeta_{\rm src}(\vec{k}_3) \bigr\rangle \\
        &= (2\pi)^3\dd\!\!\left(\sum_{i=1}^3\vec{k}_i\right) B(k_1, k_2, k_3) \,.
        \end{split}
    \end{align}
    For constant $\xi$, the bispectrum is well-approximated by\footnote{We have retained the mild scale dependence from the spectral tilt neglected in earlier works that sought to describe the overall shape rather than the slight scale dependence.} \cite{Anber:2009ua,Barnaby:2010vf,Barnaby:2011vw,Anber:2012du}
    \begin{align}
        \label{eq:bis_th}
        B(k_1, k_2, k_3) &= \frac{3}{10} (2\pi)^{5/2} \frac{\As^3}{k_1^6} \frac{1 + x_2^3 + x_3^3}{x_2^3 x_3^3} \nonumber \\
        & \quad \times e^{6\pi|\xi|} f_3(\xi, x_2, x_3) \prod_{i=1}^{3} \left(\frac{k_i}{\kp}\right)^{(\ns-1)/2},
    \end{align}
    where $x_i = k_i/k_1$ and $f_3(\xi, x_2, x_3)$ is a shape-dependent function defined in Ref.~\cite{Barnaby:2011vw}. The bispectrum peaks for equilateral configurations, $x_1 = x_2 = 1$. In the range $2 < |\xi| < 3$, the equilateral $f_3$ is approximately \cite{Barnaby:2011vw}
    \begin{equation}
        \label{eq:f3}
        f_3(\xi, 1, 1) \simeq \frac{7.4 \times 10^{-8}}{|\xi|^{8.1}} \,.
    \end{equation}
    Similar to the sourced power spectrum, deviations from constant $\xi(\tau)$ induce a tilt in the bispectrum. An increasing $|\xi(\tau)|$ results in a blue-tilted bispectrum with greater signal-to-noise ratios expected on small scales.
	
    In addition to the sourced modifications to the power spectrum and the equilateral-peaked bispectrum, the sourced curvature perturbations generate parity-even and parity-odd parts of the primordial scalar trispectrum \cite{Niu:2022fki,Fujita:2023inz}, higher order N-point statistics, and chiral gravitational waves \cite{Sorbo:2011rz,Anber:2012du}.
    The parity violation predicted by axion-U(1) inflation can only be constrained through these higher-order correlators and tensor modes \cite{Ozsoy:2021onx,Campeti:2022acx}.
    In this work, we focus on the power spectrum and bispectrum, leaving the investigation of parity violation to future work.

\subsection{Background Evolution}
    \label{ssec:background}
	
	Observables related to the axion-gauge coupling are sensitive to the background potential shape through the time dependence of $\xi(\tau)$. We aim to illustrate this within an observationally viable scenario that aligns with current CMB \cite{WMAP:2003elm,Dunkley:2010ge,Story:2012wx,Planck:2013pxb} and large-scale structure (LSS) \cite{BOSS:2012acv,DES:2021wwk,Philcox:2021kcw,DESI:2024hhd} constraints on standard cosmological parameters, specifically the observed values of $\As$ and $\ns$, with no detection of either running or primordial gravitational waves. To achieve this, we adopt a model of low-energy inflation that produces a small tensor-to-scalar ratio, with a potential flat enough to generate the observed scalar spectral tilt.
    
    In terms of slow-roll parameters,
	\begin{align}
		\epsilon_V &= \frac{1}{2 \mpl^2}\left(\frac{V_{,\phi}}{V}\right)^2 \,, \\
		\eta_V &= \frac{1}{\mpl^2} \frac{V_{,\phi\phi}}{V} \,,
	\end{align}
	the primordial scalar amplitude and spectral tilt are estimated as \cite{Lyth:1998xn}
	\begin{align}
		\label{eq:As_sr}
		\As &\simeq \frac{H^2}{8\pi^2 \mpl^2 \epsilon_{V}} \,, \\
		\label{eq:ns_sr}
		\ns &\simeq 1 - 6\epsilon_V + 2 \eta_V \,.
	\end{align}
	All time-dependent expressions on the right-hand side of these equations are evaluated when the pivot scale exits the horizon, $-\kp\tau=1$. To achieve low-energy inflation with the requisite $\As$, we need a small $\epsilon_V$. However, this value may be too small to recover the observed $\ns$ on its own. In such cases, $\ns$ is dominated by $\eta_V$, implying $|\eta_V| \gg \epsilon_V$. Under these conditions, $\epsilon_V$ evolves significantly. The absolute value of the time-dependent parameter $\xi(\tau)$, which governs the effects of the axion-gauge coupling, can be expressed as
	\begin{align}
		\label{eq:xi_eps}
		|\xi(\tau)| = \gcs\,\mpl \sqrt{\frac{\epsilon_V(\tau)}{2}} \,,
	\end{align}
	so $\xi(\tau)$ will also evolve significantly in this scenario.

    \begin{figure}
		\centering
		\includegraphics[width=1\linewidth]{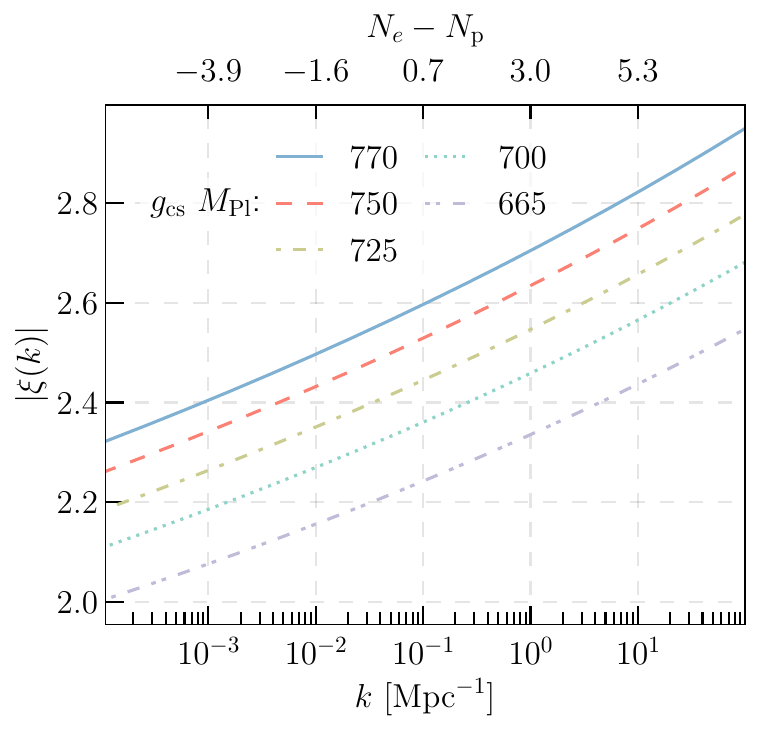}
		\caption{Absolute value of the time-dependent parameter $\xi(\tau)$ defined in Eq.~\eqref{eq:xi_eps} at horizon crossing, $-k\tau=1$, for several axion-gauge coupling strengths, $\gcs$. The top axis shows the number of $e$-folds relative to when the pivot scale exits the horizon.}
		\label{fig:xi_ks}
	\end{figure}
	
	A convenient choice satisfying these conditions is the $\alpha$-attractor potential \cite{Starobinsky:1980te,Kallosh:2013yoa}, which we parameterize as
	\begin{align}
		V(\phi) \equiv \frac{\mphi^2 \mpl^2}{2} \left(1 - e^{-\alpha_V\phi \mpl^{-1}}\right)^2 \,.
	\end{align}
	The dimensionless parameter $\alpha_V$ is often chosen such that $1.5\alpha_V^2 = 10$; we adopt this choice here. The parameter $\mphi$ has dimensions of mass and sets the height of the potential. Slow-roll inflation occurs when $\alpha_V\phi \mpl^{-1}\gg1$.
	
    In practice, we solve the background equations numerically rather than relying on the approximate expressions in Eqs.~\eqref{eq:As_sr} and \eqref{eq:ns_sr}. The background equations are
    \begin{align}
        \label{eq:bgacc}
        a'' &= \frac{a^3}{6\mpl^2}\left(\bar{\rho} - 3 \bar{P} \right) \,, \\
        \label{eq:bginf}
        \bar{\phi}'' &= -2\hub\bar{\phi}' - a^2 \langle V_{,\phi}\rangle - \gcs a^{-2} \langle \vec{E}\cdot\vec{B} \rangle \,.
    \end{align}
    The acceleration equation for the scale factor is sourced only by the inflaton's kinetic, gradient, and potential energies, as the gauge field's stress-energy tensor is traceless. The background energy density and pressure are given by
    \begin{align}
        \bar\rho &= \frac{1}{2 a^2}\bigl\langle(\phi')^2 + |\grad \phi|^2\bigr\rangle + \bigl\langle V(\phi) \rangle + \frac{1}{2 a^4} \langle E^2 + B^2 \bigr\rangle \,,\\
        \bar{P} &= \frac{1}{2 a^2}\bigl\langle(\phi')^2 - \frac{1}{3}|\grad \phi|^2\bigr\rangle - \bigl\langle V(\phi) \bigr\rangle + \frac{1}{6 a^4} \bigl\langle E^2 + B^2 \bigr\rangle \,.
    \end{align}
    To choose model parameters and set up initial conditions, we neglect the $\vec{E}\cdot\vec{B}$ term in the inflaton background equation of motion, as we are far from the strong backreaction regime during this phase.
    
    Starting from a very early time with $\bar{\phi} = 2.7~\mpl$, roughly 20 $e$-folds before the pivot scale's horizon exit, we numerically integrate Eqs.~\eqref{eq:bgacc} and \eqref{eq:bginf}. This approach ensures that we are on the attractor solution. For $\ns = 0.9665$ \cite{Planck:2018vyg}, and $\alpha_V=2.582$, we find the inflaton value at pivot scale horizon crossing $\bar{\phi}_{\rm p}=2.594~\mpl$. The derivative with respect to cosmic time, $\dee t = a \dee\tau$, is $\dot{\bar{\phi}} = \bar{\phi}'/a$. We find $\dot{\bar{\phi}}_{\rm p}=-2.717\times 10^{-3}~\mpl~\mphi$. With the scale factor normalization described below, $a_{\rm p} = 0.1226$, yielding $\bar{\phi}_{\rm p}'=-3.332\times 10^{-4}~\mpl~\mphi$.
    
    The corresponding slow-roll parameters are $\epsilon_{V}=2.345\times10^{-5}$ and $\eta_{V}=-1.766\times10^{-2}$. The tensor-to-scalar ratio from the inflaton alone is $r\simeq16\epsilon_V=3.752\times10^{-4}$, which is well within current observational bounds \cite{BICEP:2021xfz,Campeti:2022vom}. While the axion-U(1) model generates additional gravitational waves through inflaton-gauge interactions, we neglect these contributions, as they are typically negligible in the regime $2 < |\xi| < 3$ considered in this work. 
	
    After fixing the spectral tilt and, consequently, the slow-roll parameter $\epsilon_{V}$, we determine the inflationary energy scale set by $\mphi$ through the primordial scalar power spectrum amplitude. With $\As = 2.105\times10^{-9}$ \cite{Planck:2018vyg}, we find $\mphi = 4.653\times 10^{-6}~\mpl$, or $\mphi = 1.133 \times 10^{13}~{\rm GeV}$. This corresponds to an inflationary energy density of $\bar{\rho}\simeq 5.398\times10^{-12}~\mpl^{4}\simeq 1.898\times 10^{62}~\rm{GeV}^{4}$. Our simulation parameters are summarized in Table~\ref{tab:bgi}.

    Based on this inflationary background, Fig.~\ref{fig:xi_ks} shows the time dependence of $\xi(\tau)$, the parameter controlling the axion-gauge interaction. The top horizontal axis shows time as the number of $e$-folds relative to the pivot scale horizon exit time. The bottom axis shows the wavenumber that exits the horizon at that time. Modes exiting the horizon at different times are affected by different axion-gauge interaction strengths. Thus, the time dependence of $\xi(\tau)$ translates to a scale dependence, affecting the shapes of correlation functions in the axion-U(1) model. As we will demonstrate, the growth of interaction strength over time produces blue-tilted correlators.
    
    To connect our simulations with observations, we must match the comoving scales in our simulations with those at late times. Specifically, we need to identify which wavenumber corresponds to the pivot scale $\kp = 0.05~{\rm Mpc}^{-1}$ today. The comoving size of the pivot scale during inflation depends on the number of $e$-folds of expansion, $N_{\rm p}$, that have occurred since the pivot scale exited the horizon. The pivot scale exits the horizon when
    \begin{align}
        \label{eq:kp}
        e^{N_{\rm p}}\frac{H_{\rm p}}{\kp} = 1 \,.
    \end{align}
    This fixes $N_{\rm p}\simeq132.4$. Neglecting backreaction from the gauge field, there are approximately $60.5$ $e$-folds of expansion from the pivot scale's horizon exit to the end of inflation (when $\epsilon_{V}=1$), implying that about $72$ $e$-folds of expansion are split between the period from end of inflation until today and a potential strong backreaction regime during inflation. 
    
    The exact value of $N_{\rm p}$ is not fixed by observations, as it depends on the details of inflation's end and reheating. Our simulations focus on the period when modes that are observationally relevant for the CMB and LSS exit the horizon, making no assumptions about the potential's behavior beyond this window. Thus, we do not specify how the $72$ $e$-folds are distributed between a possible strong backreaction regime, reheating, and later, observationally constrained eras of cosmology. It is possible that the potential's shape avoids a strong backreaction regime. Conversely, potentials that lead to a prolonged backreaction regime and exceed our $N_{\rm p}$ budget would invalidate our model assumptions.
	
    Finally, we have the flexibility to conveniently choose the scale factor normalization. Rescaling the scale factor redefines the comoving units of length and wavenumber. As detailed in the next section, we absorb the factor $\mphi$ into the lengths and times, rendering the simulation units dimensionless:
    \begin{align}
        \Lbox = L_{i} \mphi \,,
    \end{align}
    where $L_{i}$ is the comoving length of the simulation box during inflation when the scale factor has the value $a_i$. This is related to the comoving length of the box today, $L_{0}$, by
    \begin{align}
        \frac{\Lbox k_{\rm p, i}}{\mphi} = L_0 \kp \,,
    \end{align}
    where $k_{{\rm p}, i}$ is the comoving pivot wavenumber relative to $a=a_i$, and $\kp$ is the comoving pivot wavenumber relative to the scale factor today. Evaluating this expression when the pivot scale exits the horizon yields
    \begin{align}
        \Lbox = \frac{\kp\mpc}{a_{\rm p}} \frac{\mphi}{H_{\rm p}} \frac{L_0}{\mpc} \,.
    \end{align}
    By choosing the scale factor normalization
    \begin{align}
        a_{\rm p} = \frac{H_{\rm p}}{\kp \mpc\, \mphi} \,,
    \end{align}
    the dimensionless simulation box length becomes numerically equal to the comoving box length today in units of $\mpc$. Our simulation units are effectively already in $\mpc$ units and do not require additional rescaling for comparison with CMB or LSS observables.

\begin{table}
    \centering
    \small
    \renewcommand{\arraystretch}{1.3}
    \begin{tabular}{r @{\hspace{0.4em}}| c @{\hspace{0.4em}} c @{\hspace{0.4em}} c}
        \toprule
        $\Lbox$ [$\mpc$] & $10^2$ & $10^3$ & $10^4$ \\
        \midrule
        $\Ngrid$ & $2^8$ & $\{2^6, 2^7, 2^8, 2^9\}$ & $2^8$ \\
        \midrule
        $\bar{\phi}_i~[\mpl]$ & $2.593$ & $2.607$ & $2.621$ \\
        $\dot{\bar{\phi}}_i\,[10^{-3}\mpl\, \mphi]$ & $-2.594$ & $-2.500$ & $-2.412$ \\
        $a_i$ & $7.672\times 10^{-3}$ & $7.672\times 10^{-4}$ & $7.672\times 10^{-5}$ \\
        \midrule
        $\alpha_V$ & \multicolumn{3}{c}{$2.582$} \\
        $\mphi~[10^{-6}\mpl]$ & \multicolumn{3}{c}{$4.653$} \\
        $\gcs [\mpl^{-1}]$ & \multicolumn{3}{c}{$\{665,\ 700,\ 725,\ 750,\ 770\}$} \\
        \bottomrule
    \end{tabular}
    \caption{Simulation parameters. For each box length, grid size, and coupling strength, we run 20 pairs of simulations. Each pair has a unique random seed for its initial conditions. The two simulations in a pair have initial conditions with opposite phases. The background inflaton field strength and the inflaton mass parameter are chosen to match the best-fit Planck 2018 primordial power spectrum for scalar curvature perturbations \cite{Planck:2018vyg}. The inflaton time derivatives here are with respect to cosmic time, $\dot{\bar{\phi}} = \bar{\phi}' / a$.}
    \label{tab:bgi}
\end{table}

\section{Simulations}
\label{sec:sim}
	
    Our simulations\footnote{Our code is called \emph{Adaptive Lattice Evolved Fields} (ALEF), and will be made publicly available on GitHub at \href{https://github.com/dsjamieson/alef}{https://github.com/dsjamieson/alef}.} are defined in a periodic box of length $\Lbox$. The length and comoving time units are dimensionless as we absorb $\mphi$ through $\Lbox=L\,\mphi$ and $\tau_{\rm box}=\tau\,\mphi$. Similarly, for comoving wavenumbers, $k_{\rm box}=k\,\mphi^{-1}$. The fields are defined in units of $\mpl$ so that $\phi_{\rm box}=\phi\,\mpl^{-1}$ and $A_{\rm box}{}_{\mu}=A_{\mu}\,\mpl^{-1}$. Thus, for our choice of an $\alpha$-attractor model, the inflaton-potential-dominated Hubble rate satisfies the Friedmann equation:
    \begin{align}
        \left(\frac{\dee\log a}{\dee \tau_{\rm box}}\right)^2 = \frac{a^2}{3}\Bigl(1 - e^{-\alpha_V \phi_{\rm box}} \Bigr) \,,
    \end{align}
    where every quantity is in dimensionless simulation units. Similarly, the dimensionless simulation field equations are obtained by multiplying Eqs.~\eqref{eq:ARL} and \eqref{eq:phi2} by $\mpl~{\rm Mpc}^{-2}$. From here on, we assume this has been done and drop the label ``${\rm box}$'' from all dimensionless simulation variables except $\Lbox$.
	
    \subsection{Discretization}
    \label{ssec:dis}
	
    The periodic simulation box is discretized with a comoving cubic lattice of size $\Ngrid^3$ with lattice spacing $\Delta x = \Lbox/\Ngrid$. In comoving pseudo-Cartesian coordinates, all spatial vectors are given by
    \begin{align}
        \vec{r}_{ijk} = \Delta x\left(i\vec{e}_x + j\vec{e}_y + k\vec{e}_z \right) \,.
    \end{align}
    Here, the indices $i$, $j$, and $k$ are integers, and the $\vec{e}$'s are the pseudo-Cartesian unit vectors. Periodicity requires $\vec{r}_{i'j'k'} = \vec{r}_{ijk}$ for $i' = i + n_i \Ngrid$, $j' = j + n_j \Ngrid$, and $k' = k + n_k \Ngrid$ with $n_i$, $n_j$, and $n_k$ all integers. Thus, we consider only $i,j,k\in [0, \Ngrid)$.
	
    Similarly, in Fourier space, the box has a minimum wavenumber given by the fundamental mode along any of the three pseudo-Cartesian directions:
    \begin{align}
        k_{\rm F} = \frac{2\pi}{\Lbox} \,.
    \end{align}
    All wavenumbers can be expressed as
    \begin{align}
        \vec{k}_{ijk} = k_{\rm F}\left(i\vec{e}_x + j\vec{e}_y + k\vec{e}_z \right) \,,
    \end{align}
    where again $i$, $j$, and $k$ are integers. Fourier space periodicity also requires $\vec{k}_{i'j'k'} = \vec{k}_{ijk}$ for $i' = i + n_i \Ngrid$, $j' = j + n_j \Ngrid$, and $k' = k + n_k \Ngrid$, with $n_i$, $n_j$, and $n_k$ all integers. In particular, if $i$, $j$, or $k$ are multiples of $\Ngrid$, the corresponding component of the wave vector is equivalent to zero. Also, any component at index $i>\Ngrid/2$ is equivalent to the wavenumber magnitude at index $\Ngrid - i$ pointing in the negative direction. For boxes with even $\Ngrid$, the largest wavenumber is given by the Nyquist wavenumber,
    \begin{align}
        k_{\rm Ny} = k_{\rm F} \frac{\Ngrid}{2} \,,
    \end{align}
    so we consider only pseudo-Cartesian wave vector components with $i\in(-\Ngrid/2, \Ngrid/2]$. The values $i=\pm \Ngrid/2$ are equivalent, so the Nyquist mode is neither positively nor negatively oriented and should be interpreted as a standing wave in the box. For boxes with odd $\Ngrid$, we round $\Ngrid/2$ down to the nearest integer so the corresponding Nyquist mode is not included in the box, and we have $i\in[-(\Ngrid-1)/2, (\Ngrid-1)/2]$.
	
    The fields take values
    \begin{align}
        \delta\phi_{ijk}(\tau) &= \delta\phi(\tau, \vec{r}_{ijk}) \,, \\
        A_{\rm R/L}{}_{ijk}(\tau) &= A_{\rm R/L}(\tau, \vec{r}_{ijk}) \,,
    \end{align}
    on the coordinate-space lattice. These have modes
    \begin{align}
        \delta\tilde{\phi}_{ijk}(\tau) &= \delta\tilde{\phi}(\tau, \vec{k}_{ijk}) \,, \\
        \tilde{A}_{\rm R/L}{}_{ijk}(\tau) &= \tilde{A}_{\rm R/L}(\tau, \vec{k}_{ijk}) \,,
    \end{align}
    which we distinguish from the coordinate-space field variables using tildes in this section. The modes are given by the discrete Fourier transform (DFT)
    \begin{align}
        \delta\tilde{\phi}_{ijk}(\tau) = \Delta x^3 \sum_{i,j,k} \delta\phi_{ijk} \exp(i\vec{k}_{ijk}\cdot\vec{r}_{ijk}) \,.
    \end{align}
    The sum for each component runs from 0 to $\Ngrid - 1$. The inverse DFT is defined as
    \begin{align}
        \delta\phi_{ijk}(\tau) = \left(\frac{k_{\rm F}}{2\pi}\right)^3 \sum_{i,j,k} \delta\tilde{\phi}_{ijk} \exp(-i\vec{k}_{ijk}\cdot\vec{r}_{ijk}) \,.
    \end{align}
    Here, the sums are from $-\Ngrid/2+1$ to $\Ngrid/2$ if $\Ngrid$ is even, or from $-(\Ngrid-1)/2$ to $(\Ngrid-1)/2$ if it is odd. With this discretization scheme, spatial derivatives are conveniently computed using the pseudospectral method, so under DFT
    \begin{align}
        \grad \delta\phi_{ijk}(\tau) \rightarrow i \vec{k}_{ijk} \delta\tilde{\phi}_{ijk}(\tau) \,,
    \end{align}
    with no summation over repeated indices. 
    
    Some care should be taken for even $\Ngrid$ when treating the Nyquist modes. Since the Nyquist modes are neither positively nor negatively oriented, their contributions to spatial gradients are set to zero. However, for pseudospectral Laplacians, under DFT
    \begin{align}
        \grad^2 \delta\phi_{ijk}(\tau) \rightarrow - k_{ijk}^2 \delta\tilde{\phi}_{ijk}(\tau) \,,
    \end{align}
    with $k_{ijk}^2 = \vec{k}_{ijk} \cdot \vec{k}_{ijk}$. For the Laplacian, there is no issue with the orientation of the Nyquist mode, so it is retained. This has the awkward consequence that pseudospectral Laplacians are not the same as the divergence of a pseudospectral gradient. Nevertheless, this is consistent with interpreting the Nyquist mode as a standing wave, which has equal positive and negative contributions that cancel in the gradient but contribute to the Laplacian of a scalar field.
	
    Throughout the rest of this paper, we drop the $ijk$ subscript labels on the field variables and instead denote them with wave vector or position vector arguments, with the understanding that these vectors are restricted to their discrete grid values. We also return to distinguishing between coordinate space fields and their Fourier modes using the arguments of the fields rather than placing tildes over the field variables.

    \subsection{Nonlinear interactions}
    \label{ssec:nl}
	
    The nonlinear terms of the sources defined in Eqs.~\eqref{eq:sARL}, \eqref{eq:sphi}, and \eqref{app:eq:svarphi} correspond to convolutions of modes in Fourier space. These would be expensive to compute directly, but they correspond to local operations in position space. Thus, to evaluate the nonlinear sources, we first compute the inverse DFT of the fields, evaluate the position space sources, and then take the DFT of the sources to obtain Fourier space source modes.
	
    Applying this procedure directly to the fields leads to spurious, resolution-dependent effects that are most noticeable on large scales. These spurious effects are due to the large amplitudes of small-scale vacuum fluctuations, which is the same cutoff dependence observed in perturbative nonlinear field theory without renormalization. A rigorous treatment requires properly renormalizing nonlinear local field operators \cite{Ballardini:2019rqh}, accounting for the effects of sub-Nyquist modes that are not included in the simulation box. In this way, we would solve the field equations with nonlinear field products renormalized at the cutoff scale defined by our lattice spacing, and the simulations would converge when increasing the resolution.

    Instead of pursuing this more rigorous approach, we rely on the intuition that the small-scale modes are vacuum fluctuations that, after renormalization, should have negligibly small contributions to the nonlinearity affecting modes exiting the horizon at any given time. We can then filter out the small-scale modes using a Gaussian smoothing kernel before evaluating any nonlinear terms in the equations of motion. We define separate smoothing kernels for the inflaton and gauge fields.
    \begin{align}
        \phi^{\rm s}(\tau, \vec{k}|\lambda_\phi) &= \phi(\tau, \vec{k}) \exp\left(-\frac{\lambda_\phi^2 k^2}{2}\right) \,, \\
        A_{\rm R/L}^{\rm s}(\tau, \vec{k}|\lambda_A) &= A_{\rm R/L}(\tau, \vec{k}) \exp\left(-\frac{\lambda_A^2 k^2}{2}\right) \,,
    \end{align}
    where the superscript $s$ indicates a \emph{smoothed} field. The modes of the nonlinear sources in Eqs.~\eqref{eq:sARL}, \eqref{eq:sphi}, and \eqref{app:eq:svarphi} are evaluated as
    \begin{align}
        \label{eq:source_phi_scalar}
        S^{\varphi}(\tau, \vec{k}) &= \gcs \mathcal{FT}\Bigl[ \mathcal{FT}^{-1}\left[i\vec{k} \delta\phi^s\right] \cdot \mathcal{FT}^{-1}\left[\vec{B}^s\right]\Bigr],
        \\[1em]
        \label{eq:source_A_vector}
        \vec{S}^{\vec{A}}(\tau, \vec{k}) &= -\gcs \mathcal{FT}\Bigl[ \mathcal{FT}^{-1}\left[\delta\phi^s{}'\right] \mathcal{FT}^{-1}\left[\vec{B}^s \right]
        \nonumber \\
        & \quad {} + \mathcal{FT}^{-1}\left[i\vec{k}\phi^s\right] \times \mathcal{FT}^{-1}\left[\vec{E}^s\right] \Bigr],
        \\[1em]
        \label{eq:source_phi_scalar_2}
        S^{\phi}(\tau, \vec{k}) &= - a^2\mathcal{FT}\Bigl[V_{,\phi}\left(\mathcal{FT}^{-1}\left[\phi^s\right]\right)\Bigr] - a^2 V_{,\phi\phi}\left(\bar{\phi}\right)\delta\phi^s
        \nonumber \\
        & \quad - \gcs a^{-2} \mathcal{FT}\Bigl[\mathcal{FT}^{-1}\left[\vec{E}^s\right]\cdot\mathcal{FT}^{-1}\left[\vec{B}^s\right]\Bigr] \,.
    \end{align}
    The equation for $S^{\phi}$ neglects the zero mode, which we set to zero since we deal with the background separately. Nonlinear terms in the background equations for $a(\tau)$ and $\bar{\phi}(\tau)$, Eqs.~\eqref{eq:bgacc} and \eqref{eq:bginf}, are similarly evaluated with the smoothed fields $\phi^s$ and $A_{\rm R/L}^s$ to control spurious contributions from small-scale vacuum fluctuations.
	
    We choose different smoothing scales for the inflaton and gauge fields because their nonlinearities become important at different scales. For the gauge field, the nonlinearities are sourced predominantly by wavenumbers in the range given in Eq.~\eqref{eq:xi_bounds}, where coupling to the time-dependent inflaton background amplifies the linear mode functions. We thus choose a time-dependent smoothing scale
    \begin{align}
        \label{eq:lambdaA}
        \lambda_{A}(\tau) = \frac{0.05}{|\xi(\tau)| \hub(\tau)} \,,
    \end{align}
    smoothly cutting off the vacuum gauge fluctuations that should have a negligible impact. For the inflaton, we use the time-dependent horizon to define the smoothing scale
    \begin{align}
        \label{eq:lambdaphi}
        \lambda_\phi(\tau) = \frac{0.05}{\hub(\tau)} \,.
    \end{align}
    
    In Appendix \ref{app:conv}, we show how the simulation results change when varying these smoothing scales. If the smoothing scales are too large, the effects of nonlinearity are systematically underestimated and noticeably damped. If they are too small, spurious nonlinear couplings affect the shapes of correlation functions on large scales. Our choice sits comfortably between these extreme cases, where the spurious effects of vacuum fluctuations are smoothed away while the dominant nonlinear effects of modes approaching the horizon are, to a good approximation, retained. Our results are stable at the percent level under variation of these smoothing scales by $50\%$.

    \subsection{Comparison to finite difference schemes}
    \label{ssec:fds}
    Lattice inflation codes often take an alternative approach to the one presented here, solving the equations of motion in coordinate space rather than Fourier space \cite{Felder:2000hq,Figueroa:2021yhd,Caravano:2025klk}. These coordinate space simulations estimate spatial derivatives using finite difference schemes, which introduce errors in the dispersion relation and can lead to inaccurate phase evolution (see, for example, Ref.~\cite{Caravano:2021pgc}). Additionally, controlling the spurious UV effects from nonlinear couplings is difficult to achieve in coordinate space simulations, where smoothing is a convolutional operation. Simulating in Fourier space thus has several important advantages for numerical accuracy and nonlinear convergence.
    
    The tradeoff is that Fourier space codes must resort to DFTs in order to evaluate nonlinear couplings. Since DFTs admit efficient and highly optimized $\mathcal{O}(N\log N)$ algorithms, this cost becomes significant only when the grid size necessitates distributing memory across multiple compute nodes. In this case, the internode communication, while highly optimized, incurs a noticeable cost. The parallel efficiency of pseudospectral codes decreases with additional compute nodes due to communication overhead in the distributed DFT, although the algorithm scales as $\mathcal{O}(N\log N)$ and is thus in absolute terms efficient.
    
    In practice, scaling is limited by the available CPU memory, which is typically not an issue, as modern high-performance compute clusters supply 512~GB to 1~TB of memory per node. Accelerating the simulation algorithm presented here on GPUs is currently limited only by the significantly lower amounts of memory available on GPU nodes. The increase of GPU memory in the near future will present an opportunity for significantly faster inflation simulations.
    
    \subsection{Initial Conditions and Grid Refinement}
    \label{ssec:ic}
	
    We initialize the simulations 4 $e$-folds before the fundamental mode exits the horizon (see Table~\ref{tab:bgi}). This start time is sufficiently early for nonlinear effects to imprint on the large-scale modes, but not so early that we spend excessive computational resources simulating trivial, linear dynamics. In Appendix \ref{app:conv}, we demonstrate that our simulations converge when starting at any time earlier than 3 $e$-folds before the fundamental mode's horizon exit.
	
    The initial field fluctuations $\delta\phi(\tau_i,\vec{k})$ and $A_{\rm R/L}(\tau_i,\vec{k})$ are drawn as Gaussian random fields from the linear mode functions, which are the solutions to Eqs.~\eqref{eq:ARL} and \eqref{eq:phi2} with the nonlinear sources (including $\vec{E}_{\parallel}$) set to zero. The mode functions are chosen to satisfy the Bunch-Davies vacuum conditions, so the field configuration is drawn from the vacuum in the asymptotic past and linearly rescaled to the start of the simulation. We numerically solve the linearized equations starting 20 $e$-folds before the start of the simulation, where all modes are well described as plane-wave vacuum fluctuations. In configuration space, the field fluctuations are real valued, which means their modes are Hermitian,
    \begin{align}
        \delta\phi(\tau_i,-\vec{k}) &= \delta\phi^*(\tau_i,\vec{k}) \,, \\
        A_{\rm R/L}(\tau_i,-\vec{k}) &= A_{\rm R/L}^*(\tau_i,\vec{k}) \,.
    \end{align}
    We enforce this reality condition after independently initializing all field modes.
	
    Since we are smoothing the nonlinear terms, modes much smaller than the smoothing scale have negligible impact on the nonlinear dynamics. These modes oscillate rapidly, limiting the time stepping and causing an increase in computational cost due to the increased number of modes on small scales. If we initialize all the modes simultaneously, we expend a significant amount of computational power simulating linear dynamics of vacuum fluctuations. Instead, we implement a temporal grid refinement scheme, injecting new small-scale modes when the Nyquist mode of the current resolution is 4 $e$-folds from horizon exit. Initializing small-scale modes at later times significantly accelerates the simulations, allowing much larger time steps at earlier stages of the simulation. We have checked that running with and without the temporal mesh refinement has a negligible impact on our results.
	
    The newly injected modes are initialized using the linear mode functions interpolated at their injection times. We use the same random seed in an algorithm that systematically initializes the modes from large scales to small scales, so the initial conditions of the later-injected modes are the same as they would be if all modes were initialized at the beginning of the simulation, ensuring reproducibility.
	
    We begin with a grid size of $\Ngrid = 16$, and increase by factors of 2, halving $\Delta x$, until we reach the final desired resolution of $\Ngrid = 256$. To check convergence, we also run simulations with final resolutions of $\Ngrid = 64$, $128$, and $512$. The convergence tests are presented in Appendix \ref{app:conv}.
	
    \subsection{Time integration}
    \label{ssec:ti}
	
    Due to the derivative coupling between the axion and gauge fields, the Hamilton equations of motion for the fields and their conjugate momenta do not have a separable form. For such systems, symplectic integrators such as leapfrog algorithms, which have convenient energy-conserving properties, are inefficient. The conjugate momenta depend on the field derivatives, so kick and drift operators cannot be evaluated independently. Since this class of numerical integrators is inefficient for the axion-U(1) model, we instead opt for a high-order embedded Runge-Kutta integration scheme.
	
    We use a 7(6) order integrator that simultaneously evaluates the 7th- and 6th-order Runge-Kutta scheme \cite{Butcher2016}. We then use the difference between the two schemes to estimate the integration error. We estimate errors for both the real and imaginary parts of all field modes and their derivatives, as well as for background quantities.
    
    Setting a relative error tolerance of $\epsilon_{\rm rel}=10^{-8}$ and an absolute error tolerance of $\epsilon_{\rm abs}=10^{-12}$, we reject time steps that exceed these error thresholds and reduce the step size. If a step is accepted, the step size is increased or decreased to aim for an error that is 90\% below the threshold in the next time step. This dynamical time stepping controls the numerical integration error and automatically adjusts the time step according to the fastest-varying quantity in the simulation, while minimizing the number of rejected steps. At early times, the fastest-varying quantities are Nyquist modes of the gauge field time derivative, which are rapidly oscillating vacuum fluctuations. As the inflaton fluctuations cross the horizon, their time derivatives rapidly fall off, becoming the fastest-varying quantities and limiting the step size.
	
    We integrate the modes in our simulation box until 20 $e$-folds after the Nyquist mode has exited the horizon. At this point, the modes have frozen out and are well-converged with respect to the end time of the simulation. The primordial curvature perturbations are linearly estimated from Eq.~\eqref{eq:zeta}, although the conversion between the inflaton and scalar curvature fluctuations is in principle nonlinear. The nonlinear conversion can be obtained, using the $\delta N$ formalism, as the fluctuations in the number of $e$-folds needed to evolve each lattice site to a constant inflaton hypersurface \cite{Salopek:1990jq,Sasaki:1995aw,Sugiyama:2012tj,Caravano:2025diq,Caravano:2025klk}. We have verified that the difference between the linear and nonlinear conversions is negligible, as expected given the small amplitudes of the perturbations.
    
    \subsection{Paired phase-reverse simulations}
    \label{ssec:pairs}

    \begin{figure*}
        \centering       
        \includegraphics[width=\linewidth]{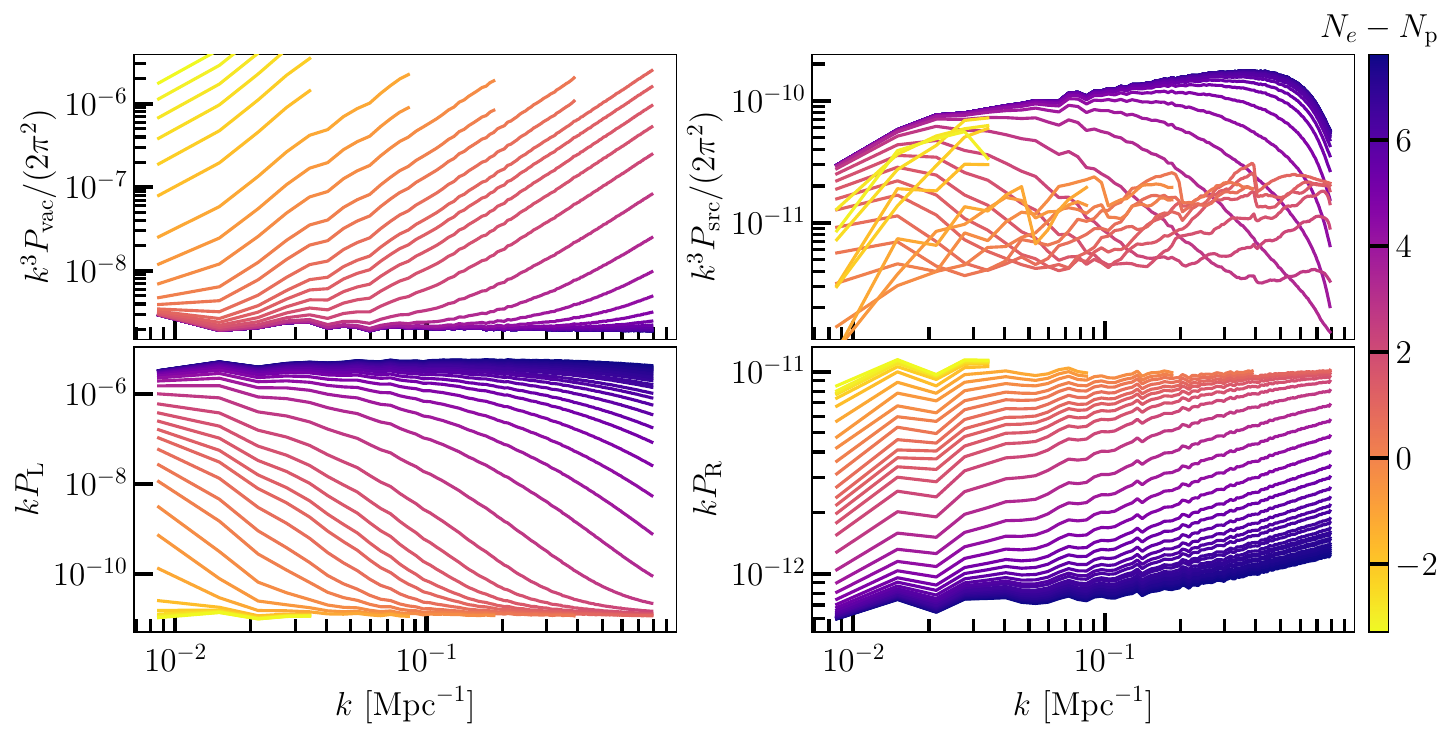}
        \caption{Time evolution of the power spectra of the vacuum curvature perturbations (top-left), the sourced curvature perturbations (top-right), the left-handed gauge modes (bottom-left), and right-handed gauge modes (bottom-right). The color bar indicates time as the number of $e$-folds relative to when the pivot scale exits the horizon. These power spectra were measured from a single pair of simulations with $\gcs=750~\mpl^{-1}$, $\Lbox=10^3~\mpc$, and $\Ngrid=256$.}
        \label{fig:time_series}        
    \end{figure*}
    
    We run pairs of simulations with the same random seed but with opposite phases. From Eq.~\eqref{eq:vac_src}, the vacuum fluctuations have opposite signs in coordinate space, while the leading sourced fluctuations, which are quadratic in the vacuum fluctuations, have the same sign. We have
    \begin{align}
        \zeta_{\pm}(\vec{k}) = \pm\zeta_{\rm vac}(\vec{k}) + \zeta_{\rm src}(\vec{k}) \,,
    \end{align}
    for the original simulation ($+$) and its phase-reversed pair ($-$). The pair will also have nearly identical integration errors. We isolate the vacuum and sourced fluctuations:
    \begin{align}
        \label{eq:zetavs}
        \zeta_{\rm vac/src}(\vec{k}) = \frac{1}{2}\bigl(\zeta_{+}(\vec{k}) \mp \zeta_{-}(\vec{k})\bigr) \,.
    \end{align}

    For the scales and couplings we consider, the primordial curvature power spectrum is dominated by the vacuum autopower spectrum on large scales, while the sourced part contributes only on small scales. Eventually, the sourced autopower spectrum would dominate on very small scales, but these are far smaller than the modes we simulate and the modes we can reliably use for cosmological inference with observations. The cross power spectrum between the vacuum and sourced curvature perturbations is negligibly small for the modes we simulate.

\section{Results}
\label{sec:res}

  \begin{figure*}
        \centering
        \includegraphics[width=0.65\linewidth]{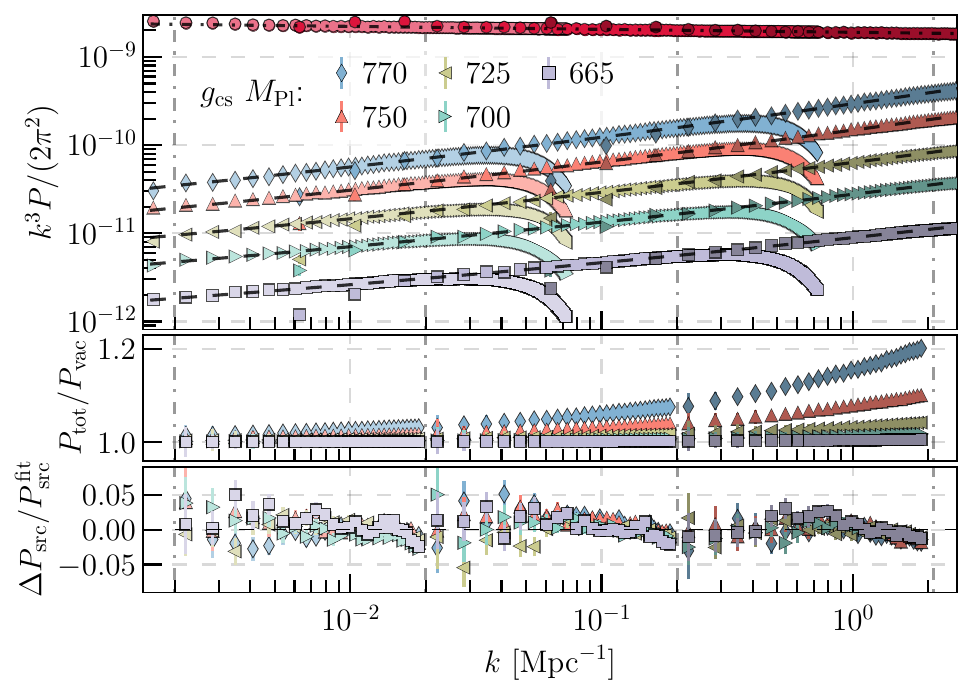}
        \caption{Power spectra from simulations with fixed $\Ngrid=256$ and different box lengths and axion-gauge coupling strengths. In the upper panel, the red data points at the top show the unsourced vacuum power, which agrees with the target primordial power spectrum (black dot-dashed line). The other data points show the sourced power spectra. All power spectra are shown for three box lengths: $\Lbox=10^4~\mpc$ (leftmost, light color), $10^3~\mpc$ (middle), and $10^2~\mpc$ (rightmost, dark color). The dashed lines show the fitting function from Eq.~\eqref{eq:pkfit}, jointly fit to all the simulated sourced power spectra with scale cuts (the vertical gray dot-dashed lines) described in the main text. The middle panel displays the ratio of total power to vacuum power, illustrating the small-scale enhancement from the axion-gauge coupling. The bottom panel shows the fractional residuals to the fitting function.}
        \label{fig:pks}
    \end{figure*}
    
    In this section, we present the main results from our simulations, including the primordial scalar curvature power spectrum and bispectrum.\footnote{The data from these simulation analyses and models we implement to fit the data are publicly available on Zenodo~\cite{jamieson_2025_17477579}.} We separate the vacuum fluctuations from the sourced fluctuations according to Eqs.~\eqref{eq:vac_src} and \eqref{eq:zetavs}. We estimate the vacuum and sourced parts of the primordial curvature power spectrum in spherical wavenumber shells. For a bin centered at wavenumber $k_i$ with width $\Delta k$, we define the spherical mode shell
    \begin{align*}
        s_i(k) =
        \begin{cases}
            1 & \text{if } |k - k_i| < \displaystyle\frac{\Delta k}{2} \\
            0 & \text{otherwise} \,,
        \end{cases}
    \end{align*}
    and define the mode shells of the vacuum and sourced curvature perturbations:
    \begin{align}
        \zeta_{{\rm vac/src}, i}(\vec{q}) = s_i(q) \zeta_{\rm vac/src}(\vec{q}) \,.
    \end{align}
    The vacuum and sourced power spectra, from Eq.~\eqref{eq:pspec_def}, are then estimated through
    \begin{align}
        P_{\rm vac/src}(k_i) = \frac{1}{\Vbox N_i} \sum_{\vec{q}} |\zeta_{{\rm vac/src}, i}(\vec{q})|^2 \,,
    \end{align}
    where $\Vbox =\Lbox^3$ is the box volume, the sum is over all modes in the box, and the normalization factor counts the number of modes in the shell:
    \begin{align}
        N_i = \sum_{\vec{q}} s_i(q) \,.
    \end{align}
	
    We also estimate the bispectrum from Eq.~\eqref{eq:bspec_def} in spherical wavenumber shells:
    \begin{align}
        \label{eq:bspec1}
        \begin{split}
            B(k_1, k_2, k_3) &= \frac{1}{\Vbox N_{123}} \\
            & \quad \sum_{\vec{q}_1,\vec{q}_2,\vec{q}_3}
            \dd\!\!\left(\sum_{i=1}^3 \vec{q}_i\right) \prod_{j=1}^{3} \zeta_{{\rm src},j}(\vec{q}_j) \,.
        \end{split}
    \end{align}
    By writing the delta function as the Fourier transform of a plane wave, the sums over $\vec{q}_i$ become inverse Fourier transforms, so the bispectrum is estimated through a sum over local products of fields in coordinate space:
    \begin{align}
        \label{eq:bspec2}
        B(k_1, k_2, k_3) = \frac{1}{\Vbox N_{123}} \sum_{\vec{x}} \zeta_{{\rm src}, 1}(\vec{x})\zeta_{{\rm src}, 2}(\vec{x})\zeta_{{\rm src}, 3}(\vec{x}) \,.
    \end{align}
    The normalization factor $N_{123}$ counts the number of closed triangles in the bispectrum bin and is computed by setting $\zeta_{{\rm src},j}(\vec{q}_j) = s_j(q_j)$ in Eq.~\eqref{eq:bspec1}.
    
    In Fig.~\ref{fig:time_series}, we show the time evolution of the power spectra for the inflaton and gauge field throughout one simulation with $\gcs=750~\mpl^{-1}$, $\Lbox=10^{3}~\mpc$, and $\Ngrid=256$. The inflaton vacuum fluctuations in the top-left panel decrease until they freeze out after crossing the horizon. The times of mesh refinement and mode injection happen each time the power spectra extend to higher wave numbers. The left-handed gauge power spectra appear in the bottom-left panel. These are exponentially enhanced and grow throughout the simulation. The right-handed gauge power spectra, in the bottom-right panel, are suppressed and negligible compared with the enhanced, left-handed modes.
    
    The physical electric and magnetic fields are suppressed by additional factors of $a^{-2}$, so their power spectra decay away outside of the horizon as $a^{-4}$, as expected for radiation perturbations. The gauge field's effects are imprinted on the sourced curvature perturbations, as demonstrated by the power spectra shown in the top-right panel of Fig.~\ref{fig:time_series}, which grow until freezing out after crossing the horizon. These results are qualitatively consistent with results from previous studies in Refs.~\cite{Caravano:2022epk,Figueroa:2024rkr}, but differ quantitatively due to a different choice of inflaton potential.

\subsection{Curvature Power Spectrum}
    \label{ssec:pk}
	
    The vacuum and sourced parts of the binned power spectra, with $\Delta k = k_{\rm F}$, are shown in Fig.~\ref{fig:pks} for simulations with the three different box volumes. We have averaged over 20 pairs of realizations, using the pairs to isolate the vacuum and sourced parts of the primordial curvature perturbation according to Eq.~\eqref{eq:zetavs}. The error bars are estimated as the standard deviation of the mean among the realizations. Each pair has independent initial conditions drawn from a unique random seed for all box lengths and coupling strengths.
    
    The vacuum power spectrum is well converged on all scales and agrees with the target primordial curvature power spectrum, demonstrating that we have correctly chosen our model parameters and accurately solved the linear parts of the inflaton field equation.
	
    The sourced power spectra are well converged for a range of scales that excludes the largest and smallest wavenumbers in the boxes. On large scales, the power is reduced due to missing nonlinear mode couplings from $k<k_{\rm F}$. On small scales, the power is similarly reduced due to missing nonlinear couplings to modes with $k>k_{\rm Ny}$. The missing mode couplings affect a wider range of small-scale modes than large-scale modes. 
    For each box length, the well-converged regions are $0.002$--$0.02~\mpc^{-1}$ ($\Lbox = 10^{4}~\mpc$), $0.02$--$0.2~\mpc^{-1}$ ($\Lbox = 10^{3}~\mpc$), and $0.2$--$2~\mpc^{-1}$ ($\Lbox = 10^{2}~\mpc$). We indicate these scales as the vertical gray dot-dashed lines in Fig.~\ref{fig:pks}. 

    \begin{figure}
		\centering
		\includegraphics[width=0.75\linewidth]{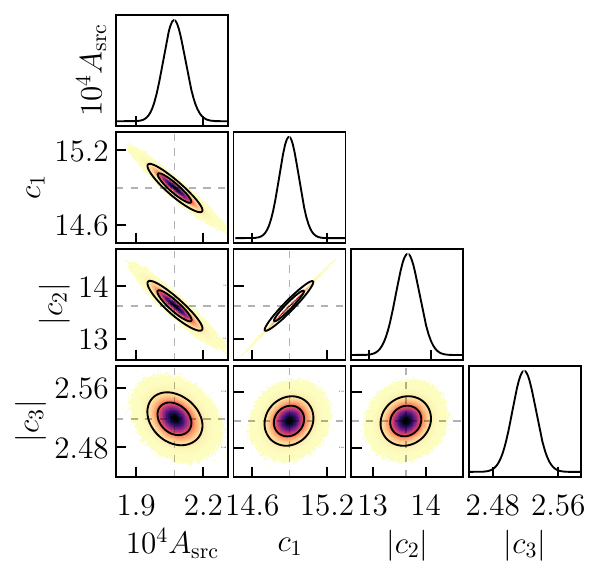}
		\caption{2D contours showing the parameters estimated from the joint fit of the sourced power spectrum model in Eq.~\eqref{eq:pkfit} to simulation data with $\Ngrid=256$, box lengths $\Lbox\, \mpc^{-1} \in \{10^4, 10^3, 10^2\}$, and axion-gauge coupling strengths $\gcs\,\mpl \in \{665, 700, 725, 750, 770\}$.}
		\label{fig:pk_corner}
	\end{figure}

    The total primordial power spectrum is the sum of the vacuum and sourced power spectra because the cross-correlation between the vacuum and sourced fluctuations is negligibly small. The sourced power is blue-tilted with significant running, which would manifest as an enhancement of total small-scale power, as illustrated in the middle panel of Fig.~\ref{fig:pks}. This enhancement differs from the typical running of $\ns$, which also modifies the large-scale power. Such a blue-tilted second component to the primordial power spectrum could explain the high-$\ell$ enhancement in CMB power reported by the ACT Collaboration \cite{ACT:2025fju}.

    \begin{table}
    	\begin{tabular}{r | @{\hspace{0.5em}} c} 
    		\toprule
    		$10^4A_{\rm src}$ & $2.071(95)$ \\
    		$c_1$ & $14.90(16)$ \\
    		$c_2$ & $-13.63(37)$ \\
    		$c_3$ & $-2.518(28)$ \\
    		\midrule
    		$\chi^2/{\rm d.o.f}$ & 1.34 \\
    		\bottomrule
    	\end{tabular}
            \caption{Best-fit parameters for the sourced power spectrum model from Eq.~\eqref{eq:pkfit}. Numbers in parentheses indicate the 95\% confidence level uncertainties in the final two digits of each parameter, estimated from their one-dimensional marginalized posterior distributions.}
            \label{tab:pk_fit}
    \end{table}
	
    We model the scale dependence and axion-gauge coupling dependence of the sourced power by assuming that the overall shape is consistent with the perturbative calculation at constant $\xi$ from Eq.~\eqref{eq:ps_th}, multiplied by a factor accounting for the tilt and running due to the change in $\xi$ for each mode at horizon crossing:
    \begin{align}
        \label{eq:pkfit}
        \mathcal{P}_{\rm src}^{\rm fit}(k) &=
        A_{\rm src} \As^2 \left(\frac{k}{\kp}\right)^{\ns - 1} e^{c_1 |\xi(k)|} |\xi(k)|^{c_2} \left(\frac{\xi(k)}{\xi_{\rm p}}\right)^{c_3} \,.
    \end{align}
    Here, $\xi(k)$ is evaluated at horizon exit ($-k\tau=1$) for mode $k$, as shown in Fig.~\ref{fig:xi_ks}. The quantity $\xi_{\rm p}$ is evaluated at the pivot scale horizon exit time. The factor $\As^2 \left(k/\kp\right)^{\ns - 1}$ gives the contributions to the amplitude and tilt coming from the vacuum power. The parameter $c_1$ sets the exponential enhancement, with $c_1=4\pi$ for constant $\xi$ \cite{Barnaby:2011vw}. The parameter $c_2$ contributes to both the tilt and to the change in amplitude due to the change in coupling strength. For constant $\xi$, $c_2\simeq-5.5$ \cite{Barnaby:2011vw}. The parameter $c_3$ affects the tilt and running only if $\xi$ is time-varying. 
    
    We jointly fit the sourced power data for all coupling strengths and all box lengths, imposing scale cuts where the power is not converged. We ran Markov Chain Monte Carlo (MCMC) chains assuming a Gaussian likelihood and wide, flat priors on all parameters except the overall amplitude $A_{\rm src}$, which was given a flat prior on $\log A_{\rm src}$. The 2D contours from the fit are plotted in Fig.~\ref{fig:pk_corner} and the best-fit parameter values are listed in Table~\ref{tab:pk_fit}.

    We find that the parameter $c_1=14.90\pm0.16$ is comparable to, but higher than, the expectation of $4\pi$ for constant $\xi$. The amplitude $A_{\rm src}$ and $c_2$ differ significantly from the constant $\xi$ values, indicating that these parameters depend on the background evolution through the slow-roll parameters, or equivalently, through the shape of the inflaton potential.

    \begin{figure*}
        \centering
        \includegraphics[width=1.\linewidth]{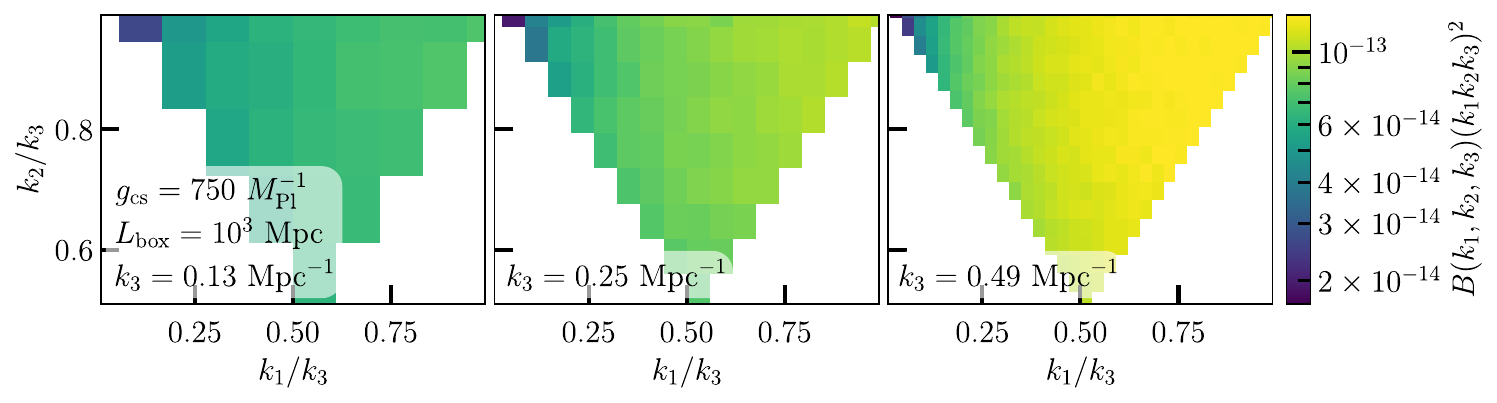}
        \caption{The shape of the sourced curvature fluctuation bispectrum for simulations with box length $\Lbox=10^3~\mpc$ and axion-gauge coupling strength $\gcs=750~\mpl^{-1}$. The bispectrum is parameterized with $k_1\leq k_2\leq k_3$, so $k_3$ sets the maximum wavenumber. The bispectrum peaks when $k_1 = k_2 = k_3$, or the upper-right corner in each of these plots.}
        \label{fig:bkmaps}
    \end{figure*}
	
    The fit is shown as the dashed line in the top panel of Fig.~\ref{fig:pks}. The fractional residuals are shown in the bottom panel of the same figure. The fit has a $\chi^2$ per degree of freedom of $1.34$, and matches the simulations well across three orders of magnitude in scale and over the full range of couplings considered. Contributions to the reduced $\chi^2$ agree among all simulation sets to within a few percent, so the fit residuals have no strong dependence on scale or coupling strength. Our simulations predict specific small-scale power modifications for axion-U(1) inflation with an $\alpha$-attractor potential. These modifications are well-described by our fitting function.

    \subsection{Bispectrum}
    \label{ssec:bk}

    We estimate the sourced bispectrum, according to Eqs.~\eqref{eq:bspec1} and \eqref{eq:bspec2}, in spherical shells of width $\Delta k = 0.0015,\ 0.015$, and $0.15~\mpc^{-1}$ for box lengths $\Lbox = 10^4$, $10^3$, and $10^2~\mpc$, respectively. We parameterize the bispectrum with three wave vector magnitudes $k_1 \leq k_2 \leq k_3$ ranging from the fundamental to the Nyquist mode in each box.

\begin{figure*}
        \centering
        \includegraphics[width=0.65\linewidth]{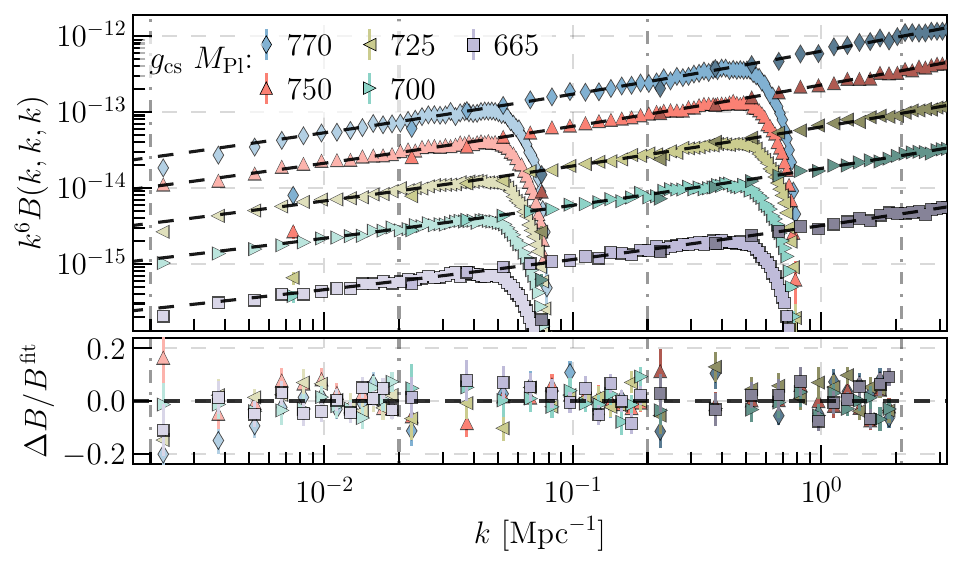}
        \caption{Equilateral configurations ($k_1=k_2=k_3$), of sourced curvature fluctuation bispectra measured from simulations with fixed $\Ngrid=256$ and different box lengths and axion-gauge coupling strengths. All bispectra are shown for three box lengths: $\Lbox=10^4~\mpc$ (leftmost, light color), $10^3~\mpc$ (middle), and $10^2~\mpc$ (rightmost, dark color). The dashed lines are the fitting function in Eq.~\eqref{eq:btemp} based on the effective power spectrum from Eq.~\eqref{eq:bfit}, jointly fit to all of the bispectra configurations (not just the equilateral ones) with scale cuts (vertical gray dot-dashed lines), as described in the main text. The bottom panel shows the fractional residuals between the bispectra and the fitting function.}
        \label{fig:bk_eq_fit}
    \end{figure*}

    We show the shape of the measured bispectrum for one box length and coupling strength value in Fig.~\ref{fig:bkmaps}. The different panels display the shape of the bispectrum at fixed $k_3$, the maximum wavenumber. The bispectrum peaks for equilateral configurations ($k_1 =k_2 = k_3$), as expected from Eq.~\eqref{eq:bis_th}, the semi-analytical perturbative analysis for the case where $\xi$ is constant \cite{Anber:2009ua,Barnaby:2010vf,Barnaby:2011vw,Anber:2012du}. In Fig.~\ref{fig:bk_eq_fit}, we show the bispectrum measured on equilateral configurations. The bispectrum increases from large scales to small scales, illustrating the blue tilt due to the time dependence of $\xi(\tau)$.
    
    The full shape of the bispectrum is not expected to have an analytical, separable form, based on the perturbative calculation from Eq.~\eqref{eq:bis_th}. Nevertheless, it is useful to have an analytical fitting function that accurately describes at least the peak of the bispectrum measured in the simulations on observationally relevant scales. The form of this fitting function is helpful in interpreting the overall bispectrum shape and could be used for simulation-based inference.

  \begin{table}
        \centering
        \begin{tabular}{l|ccc}
            \hline
            Shape, ($I$) & $b_{I}^{330}$ & $b_{I}^{222}$ & $b_{I}^{123}$ \\
            \hline
            $\text{Local}$, (loc) & 2 & 0& 0\\
            $\text{Equilateral}$, (equ) & $-6$& $-12$& 6\\
            $\text{Orthogonal}$, (ort) & $-18$ & $-48$& 18 \\
            \hline
        \end{tabular}
        \caption{Coefficients defining the bispectrum templates.}
        \label{tab:bcoefs}
    \end{table}

    We develop such a fitting function based on the standard, separable, local, equilateral, and orthogonal bispectrum shapes \cite{Komatsu:2010hc}
    \begin{align}
        \label{eq:btemp}
        \begin{split}
            B(k_1, k_2, k_3) &= f_{\rm loc} B_{\rm loc}[P_{\rm eff}](k_1,k_2,k_3) \\
            & \quad + f_{\rm equ} B_{\rm equ}[P_{\rm eff}](k_1,k_2,k_3) \\
            & \quad + f_{\rm ort} B_{\rm ort}[P_{\rm eff}](k_1,k_2,k_3).
        \end{split}
    \end{align}
    where each bispectrum depends on an underlying effective power spectrum $P_{\rm eff}(k)$, which is not necessarily related to the vacuum or sourced power spectra in a simple way. In general, even if a bispectrum is well described by such a local, equilateral, and orthogonal decomposition, it may not necessarily have the same effective power spectrum for each term. We have verified that expanding the model to include different effective power spectra for the different template shapes does not improve the fit, as the three independent effective power spectra become nearly identical for the best-fit parameter values in this expanded model.
    
    The three template shapes are all described by the form:
    \begin{align}
        \label{eq:bI}
        B_{I} =
        b_{I}^{330} B_{330}
        + b_{I}^{222} B_{222}
        + b_{I}^{123} B_{123} \,,
    \end{align}
    where the index $I$ refers to the shape (local, equilateral, or orthogonal), and the coefficients for each shape, $b_I$, are given in Table~\ref{tab:bcoefs}. The functions $B_{abc}$ are given by
    \begin{align}
        \label{eq:b330}
        B_{330}(k_1,k_2,k_3) &= P(k_1) P(k_2) + {\rm 2\ cyc.\ perms.} \,, \\
        \label{eq:b222}
        B_{222}(k_1,k_2,k_3) &= \Bigl(P(k_1) P(k_2) P(k_3)\Bigr)^{2/3} \,, \\
        \label{eq:b123}
        B_{123}(k_1,k_2,k_3) &= P(k_1)^{1/3} P(k_2)^{2/3} P(k_3) + {\rm 5\ perms.} \,.
    \end{align}
    The function $B_{333}$ has a total of three cyclically permuted terms, and the function $B_{123}$ has a total of six terms permuting $k_1$, $k_2$, and $k_3$. The sets of integers labeling each bispectrum template term denote the powers of $k_1^{-1}$, $k_2^{-1}$, and $k_3^{-1}$ that would appear for a scale-invariant power spectrum.
    
    We parameterize the effective power spectrum using a similar form to the sourced power spectrum model from the previous section:
    \begin{align}
        \label{eq:peff_fit}
        \mathcal{P}_{\rm eff}^{\rm fit}(k) = 
        \sqrt{10^{7} \As^3}\, e^{d_1 |\xi(k)|}\, |\xi(k)|^{d_2}\, \left( \frac{\xi(k)}{\xi_{\rm p}} \right)^{d_3} .
    \end{align}
    The overall normalization of the effective power spectrum is arbitrary and can be absorbed into the three $f_{I}$ parameters. The choice we make here conveniently makes $f_{\rm equ}$ of order unity. These should not be confused with the typical $f_{\rm NL}$ parameters used in standard, near-scale-invariant bispectrum analysis \cite{Komatsu:2010hc}. According to Eq.~\eqref{eq:bis_th}, there is an additional tilt factor, so the full bispectrum model is
\begin{align}
    \label{eq:bfit}
    B^{\rm fit}(k_1, k_2, k_3) &= \Bigl(
        f_{\rm loc} B_{\rm loc}[P_{\rm eff}]
        + f_{\rm equ} B_{\rm equ}[P_{\rm eff}] \nonumber \\
    & \quad + f_{\rm ort} B_{\rm ort}[P_{\rm eff}]
    \Bigr) \prod_{i=1}^{3}\left(\frac{k_i}{\kp}\right)^{(\ns-1)/2}.
\end{align}

    Our bispectrum fitting function has three parameters for the effective power spectrum shape and three for the bispectrum template coefficients, totalling six parameters. We fit these using the same MCMC method employed for the power spectrum, fitting only the scales where the sourced power spectrum is well-converged for each box volume, and jointly analyzing all coupling strengths and box lengths simultaneously. The best-fit values are given in Table~\ref{tab:bk_fit}.
        
    We find that the bispectrum shape is dominated by equilateral configurations, as expected. The orthogonal amplitude is roughly 30\% and the local amplitude roughly 10\% of the equilateral amplitude, both with negative signs. Our inferred value for parameter $d_1$ agrees remarkably well with the theoretical expectation of $3\pi\simeq9.42483$ for constant $\xi$. The fit has a $\chi^2$ per degree of freedom of 1.24, which may be overestimated due to our assumption of a diagonal covariance. 

  \begin{table}
	\begin{tabular}{r | @{\hspace{0.5em}} c} 
		\toprule
		$f_{\rm loc}$  & $-0.381(17)$ \\
		$f_{\rm equ}$ & $4.02(16)$ \\
		$f_{\rm ort}$  & $-1.247(51)$ \\
		$d_1$ & $9.365(69)$ \\
		$d_2$ & $-5.51(17)$ \\
		$d_3$ & $-1.902(12)$ \\
		\midrule
		$\chi^2/{\rm d.o.f}$ & 1.24 \\
		\bottomrule
	\end{tabular}	
        \caption{Best-fit parameters for the bispectrum model from Eqs.~\eqref{eq:btemp}--\eqref{eq:bfit}. Numbers in parentheses indicate the 95\% confidence level uncertainties in the final two digits of each parameter, estimated from their one-dimensional marginalized posterior distributions.}
        \label{tab:bk_fit}
    \end{table}

   \begin{figure}
        \centering
        \includegraphics[width=1.\linewidth]{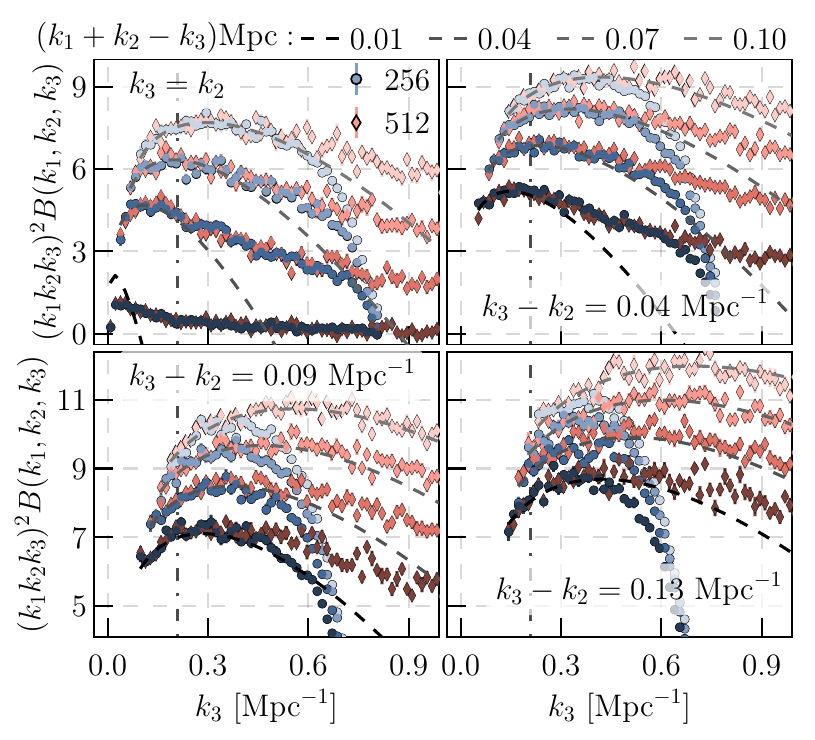}
        \caption{Bispectrum configurations as a function of the largest wavenumber, $k_3$, at fixed box length, $\Lbox = 10^3~\mpc$, and axion-gauge coupling strength, $\gcs=750~\mpl^{-1}$. Results are shown for two different grid sizes, $\Ngrid=256$ and $\Ngrid=512$. Each panel shows the bispectrum for different values of $k_3-k_2$. Each data point set displays the bispectrum at specific values of $k_1 + k_2 - k_3$ as indicated by its shade and the shade of the dashed curves, according to the legend at the top of the figure. The dashed curves show the best-fit of the model from Eqs.~\eqref{eq:btemp}--\eqref{eq:bfit}, fit to simulations with $\Ngrid=256$. The dot-dashed vertical lines indicate the scale cut used in the fit. The model continues to fit the higher-resolution simulation on scales that were not included in the fit, and where the lower-resolution simulation no longer converges.}
        \label{fig:bk_configs}
    \end{figure}

\begin{figure}
        \centering
        \includegraphics[width=1.\linewidth]{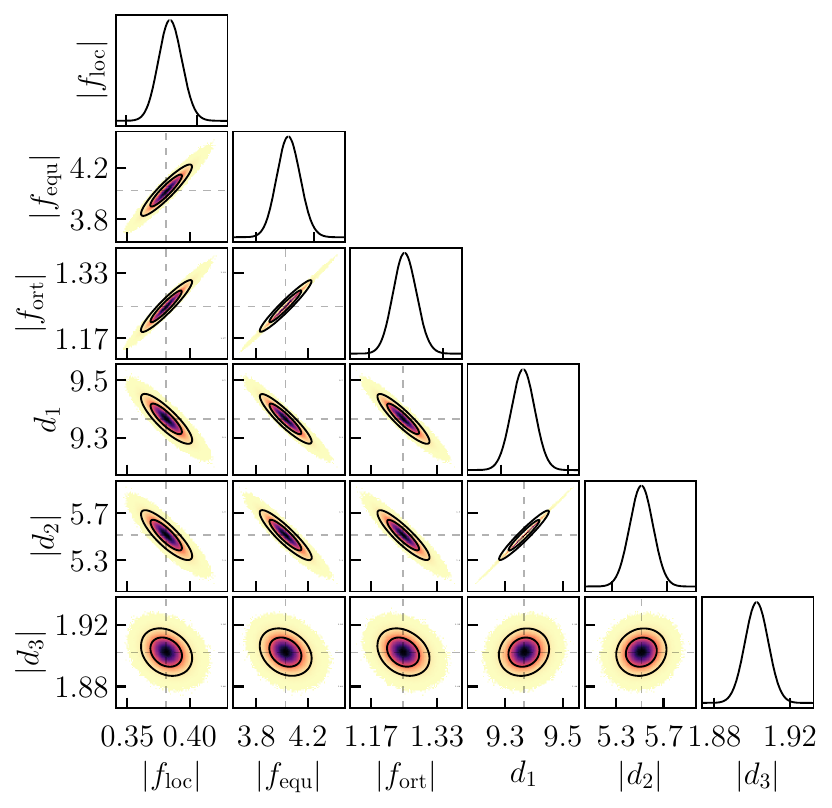}
        \caption{2D contours showing the parameters estimated from the joint fit of the bispectrum model from Eqs.~\eqref{eq:btemp}--\eqref{eq:bfit} to simulation data with $\Ngrid=256$, box lengths $\Lbox\, \mpc^{-1} \in \{10^4, 10^3, 10^2\}$, and axion-gauge coupling strengths $\gcs\,\mpl \in \{665, 700, 725, 750, 770\}$.}
    \label{fig:bk_corner}
    \end{figure}

    For the equilateral bispectrum configurations displayed in Fig.~\ref{fig:bk_eq_fit}, the fit agrees with the simulations across the full range of scales and couplings. The agreement is better than that of the sourced power spectrum model on small scales in each simulation box. This is related to the convergence of the sourced power spectrum and bispectrum, discussed in Appendix \ref{app:conv}. Fig.~\ref{fig:bk_configs} shows specific configurations, comparing simulations with $\Ngrid=256$ to high-resolution simulations with $\Ngrid=512$. The vertical dashed lines mark the scale cut from the $\Ngrid=256$ simulations, so data points to the right of these vertical lines did not contribute to the fit. Although the lower-resolution simulations lack numerical convergence on small scales, the fitting function (dashed lines) extrapolates accurately to this regime. The largest discrepancies between the fit and the data occur for highly squeezed configurations, as seen in the top left panel of Fig.~\ref{fig:bk_configs}, where the darkest data points represent the most squeezed bins. Due to their low signal-to-noise ratio, these configurations have a negligible impact on the fit and are not expected to contribute substantially to observational detection significance. The 2D contours from the fits are shown in Fig.~\ref{fig:bk_corner}.

    We find that the fitting function captures the peaks of these bispectrum curves but fails to capture the tails, indicating that these tail configurations are not accurately modelled by the separable form we have assumed. As mentioned earlier, expanding the model to include different effective power spectra for the different bispectrum template shapes does not improve the fit. This outcome supports the expectation from perturbation theory (Eq.~\eqref{eq:bis_th}) that the full bispectrum does not have a separable form. Our simulations have thus generated realizations of a non-Gaussian field with nontrivial three-point statistics that cannot be efficiently generated using the standard, template-based methods \cite{Scoccimarro:2011pz}.

    \subsection{Constraints from Observations}
    \label{sec:obs}
    
    Analyses of the CMB bispectrum have not yet searched for the specific bispectrum shape predicted by our simulations. One approach would be to implement our fitting functions, Eqs.~\eqref{eq:pkfit}, \eqref{eq:btemp}, and \eqref{eq:bfit}, which may need to be augmented to capture the full dependence on slow-roll parameters when sampling different potential shapes. Another approach would be to use the simulated fields to generate CMB maps, which may require correcting for resolution effects. In either case, the power spectrum and bispectrum should be jointly analyzed to constrain the leading-order effects of axion-U(1) inflation. We plan to implement these simulation-based inference schemes in future work.
    
    Here, we estimate the expected value of $f_{\rm NL}^{\rm equ}$ that one would detect in a conventional analysis \cite{Komatsu:2010hc,Planck:2019kim}, searching for a scale-invariant equilateral bispectrum (see Eqs.~\eqref{eq:bI}--\eqref{eq:b123} and Table~\ref{tab:bcoefs}). Assuming a diagonal, Gaussian-dominated covariance, the expected value is given by the inverse-variance-weighted sum \cite{Komatsu:2010hc}:
    \begin{align}
        f_{\rm NL}^{\rm equ} = \frac{1}{N_{\rm equ}}\sum_{k_1,k_2,k_3} \frac{B_{\rm equ}(k_1,k_2,k_3) B(k_1,k_2,k_3) s_{123}}{P(k_1)P(k_2)P(k_3)} \,,
    \end{align}
    with normalization:
    \begin{align}
        N_{\rm equ} = \sum_{k_1,k_2,k_3} \frac{\bigl(B_{\rm equ}(k_1,k_2,k_3)\bigr)^2 s_{123}}{P(k_1)P(k_2)P(k_3)} \,.
    \end{align}
    Here, $B(k_1,k_2,k_3)$ is the measured bispectrum from simulations at a fixed coupling strength. The symmetry factor, $s_{123}$, is 1 for equilateral configurations, 2 if a pair of wavenumbers are equal, and 6 if all wavenumbers are distinct. 

   \begin{figure}
        \centering
        \includegraphics[width=0.9\linewidth]{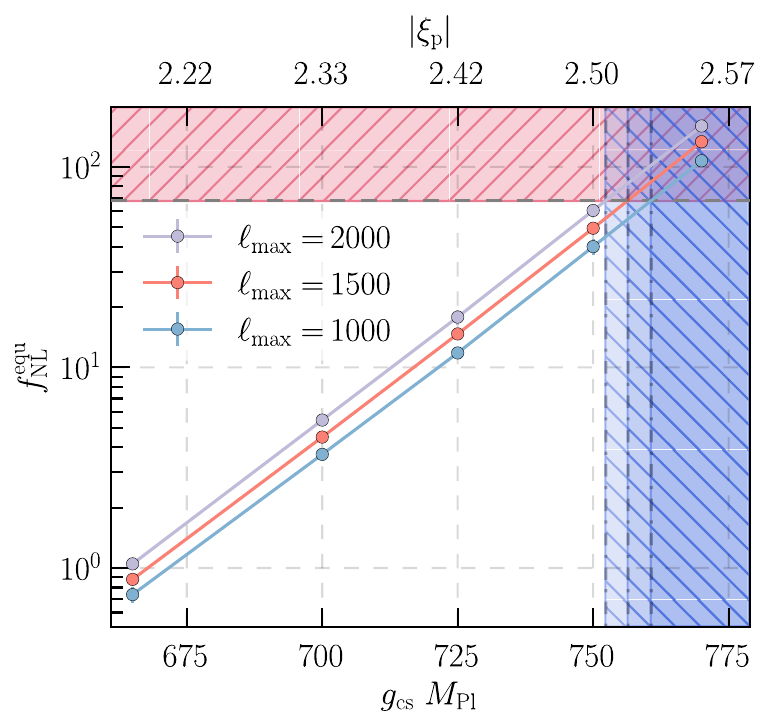}
        \caption{Constraints on $\gcs$ based on the Planck 2018 bound $f_{\rm NL}^{\rm equ} < 68$ (95\% C.L.) \cite{Planck:2019kim}, indicated by the red shaded area at the top. Values of $\gcs$ in the blue shaded areas on the right are ruled out according to the maximum $\ell$ analyzed. The top horizontal axis shows the corresponding value of $|\xi(\tau)|$, evaluated at the pivot scale horizon crossing, for the different coupling strengths.}
        \label{fig:bkcon}
    \end{figure}
    
    We take $P(k)$ to be the vacuum power spectrum and sum over all bispectrum configurations up to a $k_{\rm max}$, corresponding to
    \begin{align}
        \ell_{\rm max} = k_{\rm max} r_{\rm CMB} \,,
    \end{align}
    where $r_{\rm CMB} = 13.8~\rm{Gpc}$ is the comoving distance to the surface of last scattering. For $\ell_{\rm max} = 1000,\ 1500,$ and $2000$, we have $k_{\rm max} \simeq 0.072,\ 0.11$, and $0.14~\mpc^{-1}$, respectively. Our simulations are well-converged on these scales. 
    
    The results are plotted in Fig.~\ref{fig:bkcon}. The 2-$\sigma$ bound on positive $f_{\rm NL}^{\rm equ}$ (the axion-U(1) model predicts a purely positive bispectrum) from the Planck 2018 analysis is $f_{\rm NL} < 68$ \cite{Planck:2019kim}. From our simulations, we find constraints on the axion-gauge coupling: $|\gcs| < 760.7$, $756.4$, and $752.3~\mpl^{-1}$ for $\ell_{\rm max} = 1000,\ 1500,$ and $2000$, respectively. The sign of $\gcs$ cannot be constrained at the level of either the sourced power spectrum or bispectrum and must be determined through parity-odd observables. We can express our bounds in terms of the absolute value $|\xi_{\rm p}|$ (the time-dependent parameter $|\xi(\tau)|$ evaluated at the pivot scale horizon exit), which is labeled on the top horizontal axis of Fig.~\ref{fig:bkcon}. The constraints on $\gcs$ correspond to $|\xi_{\rm p}| < 2.535$, $2.520$, and $2.507$ for the $\alpha$-attractor axion-U(1) model we have adopted.

\section{Conclusion}
\label{sec:con}
	
    We have developed a new, high-precision code for simulating models of axion-gauge inflation. New lattice simulation techniques, described in Sec. \ref{sec:sim}, have enabled us to simulate axion-U(1) inflation with unprecedented accuracy. We ran a suite of simulations using our new code, spanning a wide range of scales and axion-gauge coupling strengths. From these simulations, we measured the power spectra and bispectra, determining their dependence on scale and coupling strength. The specific model we implemented features an $\alpha$-attractor potential, which produces significant time evolution in the axion-gauge interaction strength, as described in Subsection~\ref{ssec:background} and illustrated in Figure~\ref{fig:xi_ks}. This time evolution generates a blue tilt in the sourced spectrum and bispectrum predicted from this model, as clearly displayed in Figs.~\ref{fig:pks} and \ref{fig:bk_eq_fit}. We found that the bispectrum peaks on equilateral configurations, confirming expectations from perturbation theory.
                                             
    To facilitate interpretation and observational applications, we provide fitting functions for the sourced power spectrum and bispectrum. The power spectrum model accurately captures the scale dependence and coupling strength dependence from our simulation results. For the bispectrum, our model accurately describes configurations near the equilateral peak, but fails to capture the full shape dependence of the bispectrum tails, especially for squeezed configurations. Since the full shape of the bispectrum does not admit a separable form, standard techniques for generating primordial non-Gaussian fields \cite{Scoccimarro:2011pz} cannot produce realizations consistent with these statistics. Our simulations, therefore, establish a new method for generating realizations that capture the full nonseparable correlation structure predicted by the model. Such realizations are essential for simulation-based inference of inflationary models with sourced fluctuations.
    
    We have derived bounds on the axion-gauge coupling strength based on the Planck 2018 primordial bispectrum analysis, demonstrating a new method for constraining primordial physics with predictions using lattice simulations of inflation. Due to the blue tilt of this model's bispectrum, future CMB and LSS surveys that probe smaller scales will yield significantly tighter constraints. The precision and efficiency of our new simulation techniques broaden the scope of nonlinear inflationary physics that we can robustly constrain with future observations, opening new avenues for testing fundamental physics through cosmology.

\begin{acknowledgments}
    We thank Shaghaiegh Azyzy, Jiamin Hou, Matthew Johnson, Toshiki Kurita, S\'ebastien Renaux-Petel, Volker Springel, and Wei Xue for their helpful discussions. This work was supported in part by JSPS KAKENHI Grant Nos.~JP20H05850 and JP20H05859, and the Deutsche Forschungsgemeinschaft (DFG, German Research Foundation) under Germany's Excellence Strategy---EXC-2094---390783311. This work has also received funding from the European Union’s Horizon 2020 research and innovation programme under the Marie Skłodowska-Curie Grant Agreement No.~101007633. The Kavli IPMU is supported by World Premier International Research Center Initiative (WPI), MEXT, Japan. The work of AC was supported by the Initiative Physique des Infinis (IPI), a research training program of the Idex SUPER at Sorbonne Universit\'e.
\end{acknowledgments}

\appendix
		
\section{Gauge Field Equations}
    \label{app:geom}
	
    In this appendix, we present a derivation of the gauge field equations of motion that are solved numerically in our simulations. Varying the action in Eq.~\eqref{eq:action} with respect to the gauge potential yields the gauge field equation:
    \begin{align}
        \label{app:eq:gf1}
        \partial_\lambda \Bigl(\sqrt{-g}\bigl( F^{\mu\lambda}
        + \gcs \phi \tilde{F}^{\mu\lambda} \bigr) \Bigr) = 0 \,.
    \end{align}
    Using the identity
    \begin{align}
        \partial_\lambda \bigl(\sqrt{-g} \tilde{F}^{\mu\lambda} \bigr) = 0 \,,
    \end{align}
    the gauge field equation becomes
    \begin{align}
        \label{app:eq:gf2}
        \partial_\lambda \bigl(\sqrt{-g} F^{\mu\lambda}\bigr) = -\gcs\partial_\lambda\phi\sqrt{-g}\tilde{F}^{\mu\lambda} \,.
    \end{align}
    Using the definition of the comoving electric and magnetic fields in Eqs.~\eqref{eq:Edef} and \eqref{eq:Bdef} and their corresponding relationship with $\tilde{F}^{\mu\nu}$ from Eqs.~\eqref{eq:Bdual} and \eqref{eq:Edual}, the $\mu=0$ component becomes the Coulomb constraint, Eq.~\eqref{eq:coulomb1}, and the spatial components become the axion-U(1) Ampère-Maxwell law in Eq.~\eqref{eq:maxamp1}. Both equations take the form of a wave equation in the Lorenz gauge, defined by the condition $\partial_\mu A^\mu = 0$:
    \begin{align}
        \label{app:eq:coulomb1}
        -\square\varphi &= \gcs\grad\phi \cdot B \,, \\
        \label{app:eq:maxamp1}
        -\square\vec{A} &= \gcs\Bigl(\phi'\vec{B} + \grad\phi\times\vec{E}\Bigr) \,,
    \end{align}
    where $-\square f = f'' - \grad^2 f$ is the comoving wave operator.
	
    Although Eq.~\eqref{app:eq:coulomb1} has the appearance of a wave equation for the scalar potential, it is actually a constraint on the longitudinal part of the electric field. Moreover, due to our choice of gauge, taking the divergence of Eq.~\eqref{app:eq:maxamp1} yields the same equation as taking the conformal time derivative of Eq.~\eqref{app:eq:coulomb1}, so the longitudinal part of Eq.~\eqref{app:eq:maxamp1} is automatically satisfied if the Coulomb constraint is satisfied.
	
    To solve the constraint equation, we work in Fourier space, where the gauge vector potential has modes
    \begin{align}
        \vec{A}(\tau, \vec{k}) = \int d^3x\, e^{-i\vec{k}\cdot\vec{x}}\, \vec{A}(\tau, \vec{x}) \,.
    \end{align}
    We decompose the vector potential into longitudinal and transverse (right-/left-handed) components:
    \begin{align}
        \vec{A}(\tau, \vec{k}) &= \vec{e}_{\parallel}(\vec{k}) A_{\parallel}(\tau, \vec{k}) + \vec{A}_{\perp}(\tau, \vec{k})\,, \\
        \vec{A}_{\perp}(\tau, \vec{k}) &= \vec{e}_{\rm R}(\vec{k}) A_{\rm R}(\tau, \vec{k})
        + \vec{e}_{\rm L}(\vec{k}) A_{\rm L}(\tau, \vec{k}) \,,
    \end{align}
    and similarly for the modes of the electric and magnetic fields. Here, the longitudinal polarization is
    \begin{align}
        \vec{e}_{\parallel}(\vec{k}) = i \frac{\vec{k}}{k} \,,
    \end{align}
    and the transverse helicity basis is defined such that
    \begin{align}
        \vec{e}_{\parallel}(\vec{k}) \times \vec{e}_{\rm R/L}(\vec{k}) = \pm \vec{e}_{\rm R/L}(\vec{k}) \,.
    \end{align}
    The Lorenz gauge condition becomes
    \begin{align}
        A_{\parallel}(\tau, \vec{k}) = \frac{1}{k}\varphi'(\tau, \vec{k}) \,.
    \end{align}
    The electric and magnetic field modes are
    \begin{align}
            \begin{split}
        \vec{E}(\tau, \vec{k}) &= k^{-1}\vec{e}_{\parallel}(\vec{k}) \bigl(\varphi''(\tau,\vec{k}) - k^2 \varphi(\tau,\vec{k})\bigr) \\
        & \quad - \vec{e}_{\rm R}(\vec{k}) A'_{\rm R}(\tau, \vec{k})
        - \vec{e}_{\rm L}(\vec{k}) A'_{\rm L}(\tau, \vec{k}) \,,
            \end{split}
    \end{align}
    and
    \begin{align}
        \vec{B}(\tau, \vec{k}) =
        k\vec{e}_{\rm R}(\vec{k}) A_{\rm R}(\tau, \vec{k})
        - k\vec{e}_{\rm L}(\vec{k}) A_{\rm L}(\tau, \vec{k}) \,.
    \end{align}
    
    The longitudinal part of the electric field is equivalent to $-\square\varphi$, up to a factor of $k^{-1}$, so we can replace it with the gauge scalar source from the right-hand side of Eq.~\eqref{app:eq:coulomb1}:
    \begin{align}
        \label{app:eq:svarphi}
        S^{\varphi} = \gcs \grad \phi \cdot \vec{B} \,.
    \end{align}
    This relation involves only the propagating inflaton and helical gauge modes, so we use this source to eliminate the nonpropagating gauge field components, $\varphi$ and $A_{\parallel}$, from the equations of motion. Then the modes of the electric field are
    \begin{align}
        \begin{split}
            \vec{E}(\tau, \vec{k}) &=
            k^{-1} \vec{e}_{\parallel}(\vec{k}) S^{\varphi}(\tau, \vec{k})
            - \vec{e}_{\rm R}(\vec{k})A'_{\rm R}(\tau, \vec{k})
            \\ &
            \quad - \vec{e}_{\rm L}(\vec{k})A'_{\rm L}(\tau, \vec{k}) \,.
        \end{split}
    \end{align}
    This solves both Eq.~\eqref{app:eq:coulomb1} and the longitudinal part of Eq.~\eqref{app:eq:maxamp1}. Only the two helical gauge field equations remain to be solved. These are the transverse parts of Eq.~\eqref{app:eq:maxamp1}, given in Eq.~\eqref{eq:ARL} with the nonlinear source from Eq.~\eqref{eq:sARL}.

   \begin{figure*}
        \centering
        \includegraphics[width=0.92\linewidth]{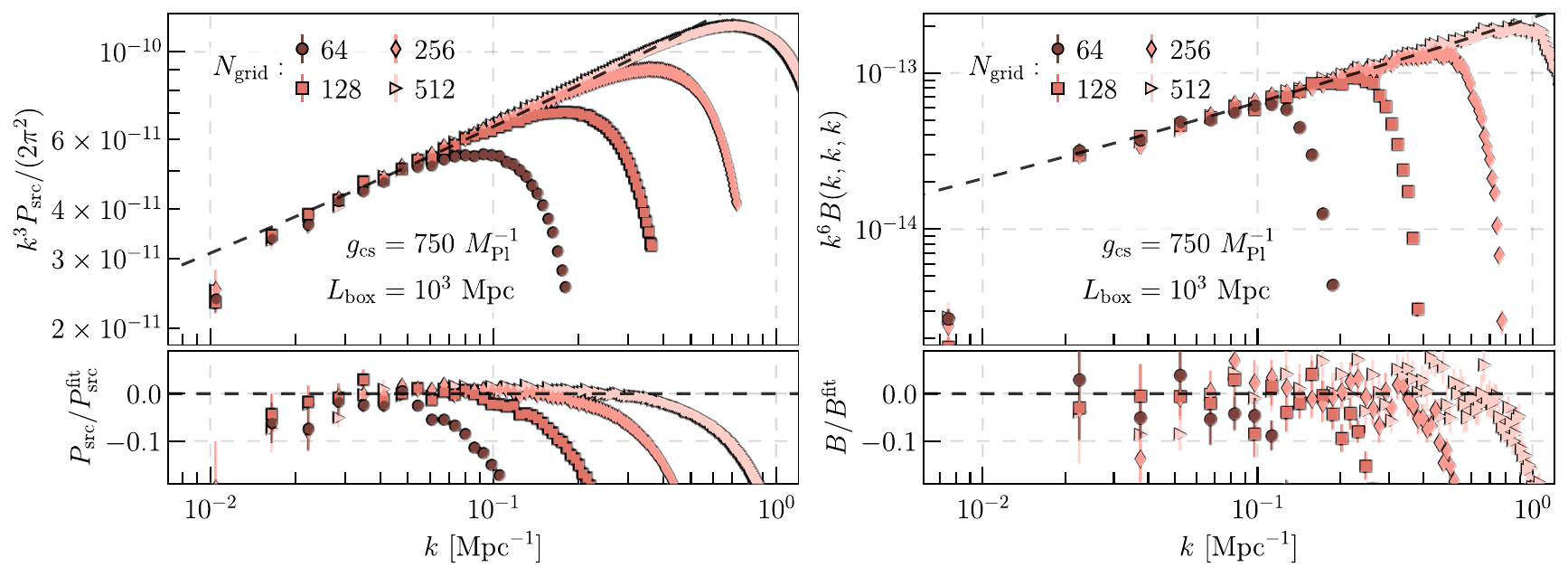}
        \caption{Convergence properties of the sourced power spectrum (left) and equilateral bispectrum (right) with respect to grid resolution $\Ngrid$. The top panels show these statistics for different grid resolutions at fixed box length, $\Lbox = 10^{3}~\mpc$, and axion-gauge coupling strength, $\gcs=750~\mpl^{-1}$. The black dashed lines show the best-fit models---Eq.~\eqref{eq:pkfit} for the power spectrum and Eqs.~\eqref{eq:btemp}--\eqref{eq:bfit} for the bispectrum---demonstrating the expected scaling when varying box length at fixed grid size. Although the largest scales are not converged with respect to varying box volume due to the absence of modes larger than the box size, they are converged with respect to increasing resolution, demonstrating that the simulations are not strongly sensitive to the small-scale cutoff imposed by the Nyquist modes. Results are averaged over the set of 20 simulations with error bars showing the standard deviation of the mean.}
        \label{fig:pk_bkeq_res}
    \end{figure*}
	
\section{Metric Perturbations in the Inflaton Field Equation}
    \label{app:seom}
    
    Here, we review the effect of linear scalar metric perturbations on the inflaton's equation of motion. Working in the spatially flat gauge, there are two scalar metric perturbations, two components of a transverse vector perturbation, and two components of the transverse-traceless gravitational wave tensor perturbation. We neglect the vector and tensor perturbations, which are sourced by nonlinear terms in the stress-energy tensor. These have negligible impact on the inflaton and gauge field equations in the perturbative regime. We consider only the scalar metric perturbations $\psi$ and $C$:
    \begin{align}
        g_{00} &= a(\tau)^2\bigl(1 + 2\psi(\tau, \vec{x})\bigr) \,, \\
        g_{0i} &= a(\tau)^2 \grad_i C(\tau, \vec{x}) \,, \\
        g_{ij} &= a(\tau)^2\delta_{ij} \,.
    \end{align}
    We eliminate these scalar perturbations to linear order in $\delta\phi$ by solving the linearized Einstein field equations algebraically.
    
    The temporal scalar perturbation is given by the divergence of the $0i$-components of the Einstein field equations:
    \begin{align}
        \label{app:eq:psi}
        \psi = -(\hub' - \hub^2)\frac{\delta\phi}{\hub\bar{\phi}'} \,,
    \end{align}
    and the 00-component of the Einstein field equations then gives
    \begin{align}
        \label{app:eq:C}
        \grad^2 C = \left(\hub' - \hub^2\right)\left[\frac{\delta\phi'}{\hub\bar{\phi}'} - \left(\frac{\bar{\phi}''}{\hub\bar{\phi}'} - \frac{\hub'}{\hub^2}\right)\frac{\delta\phi}{\bar{\phi}'}\right] \,.
    \end{align}
	
    Varying the action in Eq.~\eqref{eq:action} yields the inflaton field equation:
    \begin{align}
        \label{app:eq:phi1}
        \partial_\lambda\bigl(\sqrt{-g}\partial^\lambda\phi\bigr) = \sqrt{-g}V_{,\phi} + \gcs F_{\mu\nu}\sqrt{-g}\tilde{F}^{\mu\nu} \,.
    \end{align}
    To leading order in metric perturbations, this simplifies to Eq.~\eqref{eq:inf1}, where the second term from the right is the mass shift from the metric perturbations:
    \begin{align}
        a^2\Delta m_{\rm eff}^2\delta\phi = -\bar{\phi}'\bigl(\psi' + \grad^2 C\bigr) + 2 a^2 V_{,\phi}(\bar{\phi}) \psi \,.
    \end{align}
    Using Eqs.~\eqref{app:eq:psi} and \eqref{app:eq:C}, we eliminate the metric perturbations on the right-hand side of the above expression. Then the mass shift has the form shown in Eq.~\eqref{eq:mass_shift}. 

\section{Convergence Tests}
\label{app:conv}

    \begin{figure*}
		\centering
		\includegraphics[width=0.97\linewidth]{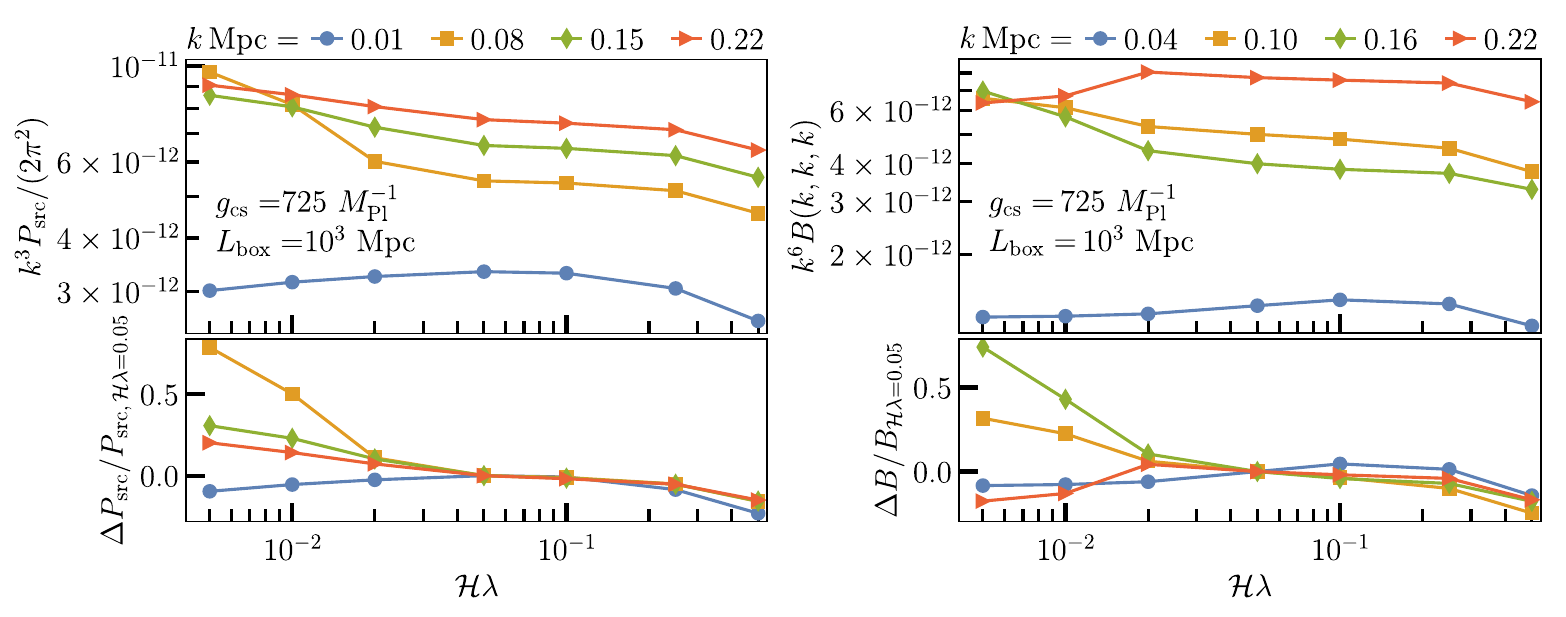}
		\caption{Convergence properties of the sourced power spectrum (left) and equilateral bispectrum (right) with respect to the smoothing scales used to regulate small-scale divergences in nonlinear interactions. The horizontal axes show the smoothing scale used for the inflaton field, $\lambda=\lambda_{\phi}$. The corresponding gauge field smoothing scales are chosen such that $\lambda_A = \lambda_{\phi} / |\xi|$. Results are displayed for a single simulation, rerun with identical initial conditions while varying $\lambda$. Undersmoothing (low $\lambda$) causes the shapes of these statistics to vary significantly due to small-scale cutoff dependence. Oversmoothing preserves the shapes but systematically lowers the amplitudes. Between these extremes lies a stable region where spectral shapes and amplitudes are insensitive to smoothing scale variations. Throughout this work, we use $\hub\lambda = 0.05$.}
		\label{fig:smoothing_conv}
	\end{figure*}
	
	\begin{figure}[ht!]
		\centering       
		\includegraphics[width=\linewidth]{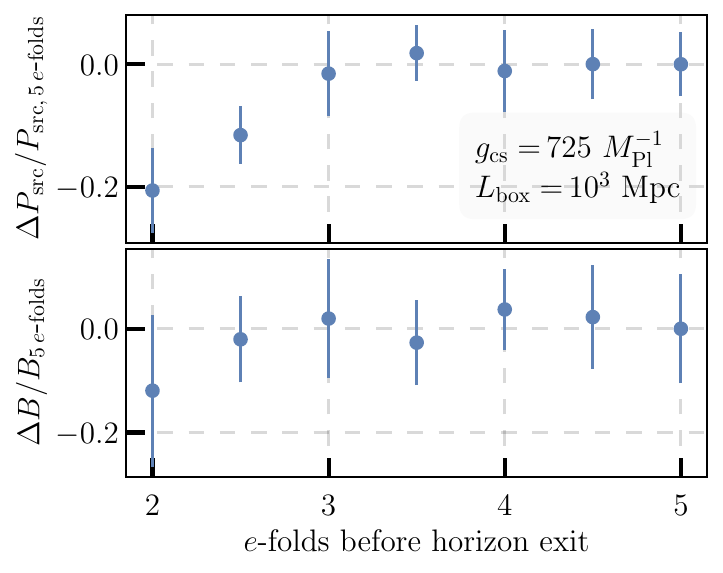}
		\caption{Convergence properties of the sourced power spectrum (top) and equilateral bispectrum (bottom) with respect to the simulation start time, quantified as the number of $e$-folds prior to the fundamental mode exiting the horizon. Results are averaged over 10 simulation pairs with error bars estimated as the standard deviation of the mean. All simulations had axion-gauge coupling strength $\gcs=725~\mpl^{-1}$, box length $\Lbox = 10^{3}~\mpc$, and $\Ngrid = 256$. Simulations initialized less than $3~e$-folds before horizon exit fail to converge. Throughout this work, we initialize modes $4~e$-folds before the largest scale exits the horizon in each resolution level.}
		\label{fig:start_times}
		
		~
		
	\end{figure}

    Here, we present convergence tests examining how our simulation results depend on resolution, smoothing scale, and start time. We tested convergence with respect to grid resolution by running a set of 20 pairs of simulations with axion-gauge coupling strength $\gcs=750~\mpl^{-1}$, box length $\Lbox = 10^{3}~\mpc$, and various grid sizes $\Ngrid\in\{64, 128, 256, 512\}$. We plot the resulting sourced power spectra and equilateral bispectrum configurations in Fig.~\ref{fig:pk_bkeq_res}. We compare these results to the maximum likelihood fitting functions from Eq.~\eqref{eq:pkfit} and Eqs.~\eqref{eq:btemp}--\eqref{eq:bfit}. All grid resolutions exhibit a lack of convergence on the largest scales due to missing mode couplings from wavelengths exceeding the box size. On small scales, each resolution deviates by more than 1\% from the converged result at about $k_{\rm Ny}/3$ for the power spectrum and at about $k_{\rm Ny}/2$ for the equilateral bispectrum configurations. The small-scale lack of convergence is due to missing sub-Nyquist modes at fixed resolution---a generic phenomenon of finite resolution in nonlinear lattice field theory simulations.

    Beyond resolution effects, we also tested convergence with respect to the Gaussian smoothing used to regulate small-scale divergences in nonlinear interactions. We ran a set of simulations setting the axion-gauge coupling strength to $\gcs=725~\mpl^{-1}$, the box length to $\Lbox = 10^{3}~\mpc$, and varying smoothing scales: $\hub\lambda \in \{0.005, 0.01, 0.02, 0.05, 0.1, 0.25, 0.5\}$. For these tests, we set $\lambda_{\phi}=\lambda$ and $\lambda_A = \lambda/\xi$, from Eqs.~\eqref{eq:lambdaA} and \eqref{eq:lambdaphi}.  All simulations had identical initial conditions. To isolate the effects of smoothing, we ran these simulations without the temporal grid refinement, initializing all modes at the start of the simulation. We plot the results in Fig.~\ref{fig:smoothing_conv}. If the smoothing is too small, spurious small-scale couplings bias the large-scale power spectrum and bispectrum configurations. As we increase the smoothing scale, the shapes of these spectra change significantly until the range $0.01 < \hub \lambda < 0.2$, where the shapes and amplitudes of the spectra stabilize against variations in the smoothing scales. At larger values of the smoothing scale, the shapes of the power spectrum and bispectrum do not change, but their overall amplitudes decrease systematically. We choose the smoothing scale $\hub\lambda=0.05$ throughout this work. With this choice, on scales where the simulations converge with respect to resolution, our results are unaffected by spurious small-scale effects due to undersmoothing, and their amplitudes are not systematically reduced due to oversmoothing.

    Finally, initializing the simulations too late biases the sourced power spectrum and bispectrum by reducing their amplitudes on large scales, missing earlier nonlinear interactions. Similarly, for the temporal grid refinement, if we inject the small-scale modes too late, we find oscillatory patterns in the correlators on the largest scales of each resolution level. We tested convergence with respect to simulation start time, varying initialization from 2 to 5 $e$-folds prior to the largest scale's horizon exit in each resolution level. We averaged the results over 20 pairs of simulations with axion-gauge coupling strength $\gcs=725~\mpl^{-1}$, box length $\Lbox = 10^{3}~\mpc$, and $\Ngrid = 256$. In Fig.~\ref{fig:start_times}, we plot the results for the largest-scale bins of the sourced power spectrum and the equilateral bispectrum configurations. Start time effects are strongest on the largest scales. The results are converged when initializing 3 $e$-folds prior to horizon exit or earlier. Throughout this work, we adopt initialization times of 4 $e$-folds prior to horizon exit, ensuring convergence across all scales of interest.

    \bibliography{main}
	
\end{document}